\documentclass[preprint,5p,times,twocolumn]{elsarticle}
\usepackage{graphics}
\usepackage{subfigure}
\usepackage{amssymb}
\usepackage{amsmath}
\usepackage{mathptmx}
\usepackage{etoolbox}
\usepackage{mathtools}
\usepackage{subcaption}
\usepackage{dcolumn}
\usepackage{booktabs}
\usepackage{hyperref}
\usepackage{multirow}
\usepackage{tabularx}
\usepackage[dvipsnames]{xcolor}

\begin{document}
\begin{frontmatter}
\title{The impact of social noise on the majority rule model across various network topologies}

\author[brin]{Roni Muslim\corref{cp}}
\ead{roni.muslim@brin.go.id}
\cortext[cp]{Corresponding author}
\author[brin,uty]{Didi Ahmad Mulya}
\ead{didiahmadmulya1@gmail.com}

\author[brin]{Zulkaida Akbar}
\ead{zulkaida.akbar@brin.go.id}
\author[ugm]{Rinto Anugraha NQZ }
\ead{rinto@ugm.ac.id}

\affiliation[brin]{
    organization={Research Center for Quantum Physics,
    National Research and Innovation Agency (BRIN)},
    city={South Tangerang},
    postcode={15314},
    country={Indonesia}
}

\affiliation[uty]{
    organization={Department of Industrial Engineering,  University of Technology Yogyakarta},
    city={Yogyakarta},
    postcode={55285},
    country={Indonesia}
}
\affiliation[ugm]{
    organization={Department of Physics,
    Universitas Gadjah Mada},
    city={Yogyakarta},
    postcode={55281},
    country={Indonesia}
}

\begin{abstract}
We explore the impact of social noise, characterized by nonconformist behavior, on the phase transition within the framework of the majority rule model. The order-disorder transition can reflect the consensus-polarization state in a social context. This study covers various network topologies, including complete graphs, two-dimensional (2-D) square lattices, three-dimensional (3-D) square lattices, and heterogeneous or complex networks such as Watts-Strogatz (W-S), Barabási-Albert (B-A), and Erdős-Rényi (E-R) networks, as well as their combinations (multilayer network). Social behavior is represented by the parameter \( p \), which indicates the probability of agents exhibiting nonconformist behavior. Our results show that the model exhibits a continuous phase transition across all networks. Through finite-size scaling analysis and evaluation of critical exponents, our results suggest that the model falls into the same universality class as the Ising model.
\end{abstract}

\end{frontmatter}
\section{\label{sec:Sec1}Introduction}
Scientists apply principles from physics to social science to understand pervasive social phenomena in society \cite{serge2016sociophysics, castellano2009statistical, galam2004sociophysics, sen2014sociophysics}. For instance, discrete opinion models, inspired by real-world dynamics, capture how individuals tend to align with majority viewpoints \cite{galam1986majority, mobilia2003majority, krapivsky2003dynamics}, how social pressures influence opinion shifts \cite{biswas2009model}, and how social validation drives conformity to prevailing trends \cite{sznajd2000opinion, sznajd2021review}. Another example is continuous opinion frameworks, which suggest that trust between individuals can influence their opinions \cite{deffuant2001mixing, weisbuch2002meet}. While most opinion dynamic models eventually lead to homogeneity, with everyone adopting the same opinion, the coexistence of minority and majority opinions can persist. To capture these complexities, scientists introduce features like social noise, which challenges prevailing opinions. The presence of such noise can lead to intriguing new phenomena worth exploring through a physics lens.

Social noise occurs when individuals or groups diverge from prevailing societal norms or expectations. This deviation often means departing from established conventions, traditions, or cultural practices, a behavior commonly known as nonconformity\cite{asch1956studies, willis1963two, willis1965conformity, macdonald2004expanding, nail2011development}.
The study of nonconformity behavior in opinion dynamics models that induce the emergence of new phases has been extensively explored across various scenarios~\cite{sznajd2011phase, nyczka2013anticonformity, javarone2014social, crokidakis2015inflexibility, chmiel2015phase, abramiuk2019independence, muslim2020phase, nowak2021discontinuous, civitarese2021external, muslim2021phase, muslim2022opinion, muslim2023effect, AZHARI2024130042, mulya2024phase}. The emergence of these novel phases sparks interest in both social sciences and physics. The phase transitions induced by nonconformity behavior are analogous to the ferromagnetic-antiferromagnetic or ferromagnetic-paramagnetic phase transitions observed in magnetic systems. In this context, these phase transitions can represent a shift from consensus to discord in socio-political discourse.

As the Ising model in physics describes, a phase transition from an ordered to a disordered phase occurs at a critical temperature. At low temperatures, spin orientations align, forming an ordered arrangement. Conversely, spin orientations become random above the critical temperature, leading to a disordered state. This concept also applies to opinion-dynamic models. During interactions and discussions, people typically agree and converge on a common viewpoint, potentially reaching a consensus. Conversely, nonconformity behavior could result in polarization or a stalemate situation.

In physics, as the Ising model exemplifies, systems with different underlying interactions or lattice structures can exhibit the same critical behavior near the critical point. This implies that observables such as critical exponents (e.g., $\beta$, $\gamma$, $\nu$) and scaling functions are independent of specific details of the system and instead depend only on the dimensionality of space and the symmetries of the order parameter \cite{chandler1987introduction}. This concept is called universality. This universality allows physicists to study critical phenomena in simpler models or theoretical frameworks and then apply their findings to understand complex systems and sociophysics.

In the social interaction model with nonconformity behavior, the network topology significantly influences the occurrence and type of phase transitions. Applying the same interaction model to different network structures can yield different results. Conversely, even with different network structures, varying degrees of connectivity, and distinct microscopic interaction mechanisms—meaning that interactions between individuals and their nearest neighbors differ across networks—they can exhibit the same critical phenomena, for example, by analyzing the model's universality. This phenomenon will be discussed in this paper.

This paper investigates how social noises, like independence and anticonformity, influence the transition from order to disorder in the majority rule model. We study this model on different types of networks to better understand the transition and the behavior of the models, specifically focusing on the universality class of the observed phase transitions. We look at various networks, homogeneous networks such as the complete graph, a two-dimensional (2-D) lattice, and a three-dimensional (3-D) lattice, as well as heterogeneous networks such as the Watts-Strogatz (W-S) network \cite{watts1998collective}, Barabasi-Albert (B-A) network \cite{albert2002statistical}, and Erdos-Renyi (E-R) network \cite{erdds1959random}. These heterogeneous networks refer to networks where nodes or edges possess diverse properties. These networks contrast with homogeneous networks, where all nodes and edges are essentially similar. Heterogeneous networks are prevalent in various real-world systems, including social networks, biological networks, transportation networks, and technological networks \cite{newman2018networks}. 

Our study demonstrates that the model defined on all networks undergoes a continuous phase transition. We also analyzed the universality class of the model by estimating the critical exponents associated with finite-size scaling analysis. Our results indicate that the model defined on the complete graph and heterogeneous networks falls into the same universality class as the mean-field Ising model. Additionally, the model on the 2-D lattice falls into the same universality class as the 2-D Ising model, and the model on the 3-D lattice also falls into the same universality class as the 3-D Ising model. Detailed results are discussed in Section \ref{sec:result}.

\section{Model and Methods}
\label{sec.2}
The majority rule model, also called the Galam majority rule model, is an opinion dynamics model that illustrates how the majority opinion in a small group (such as in a discussion) within a population always dominates or wins \cite{galam1986majority, galam2008sociophysics}. In general, the majority rule operates as follows. A small group of agents or individuals is randomly selected. Within this small group, all members interact with each other and adopt the majority opinion. This dynamic makes sense when considering groups with an odd number of members. In social psychology, when agents follow the majority opinion, they exhibit conformity behavior, often called conformist agents. Conformity means adjusting one's attitudes, beliefs, and behavior to fit the established norms of a group \cite{cialdini2004social}. In contrast to conformity, another important social behavior is nonconformity, which can be divided into two categories: anticonformity and independence. Independent behavior acts according to its own will without the influence of others. In contrast, anticonformity behavior evaluates the will of others and adopts the opposite stance~\cite{willis1963two, willis1965conformity, macdonald2004expanding, nail2011development}.

Another version of the majority model, the majority vote model with a noise parameter, where only a single opinion or spin changes based on its nearest neighbors, has been discussed in Refs.~\cite{de1992isotropic,de1993nonequilibrium,kim2021majority,campos2003small,pereira2005majority,chen2015critical}. The fundamental difference from the majority rule lies in the agents that change: in the majority rule, all agents have the potential to change based on the local majority among them, while in the majority vote model, only the central agent changes based on the majority opinion of its neighbors. In the majority vote model, it has been shown that the model, defined on various networks such as lattices and complex networks, undergoes a continuous phase transition at a specific noise parameter value and shares the same universality class as the Ising model.

In this paper, we explore the majority rule model that involves the nonconformist behavior, represented by a probability denoted as \( p \), which represents the likelihood of agents adopting either independence (model with independence) or anticonformity (model with anticonformity). Simply put, with a probability of \( p \), agents opt to act independently or display anticonformity. Conversely, with a probability of \( 1-p \), agents conform by aligning with the majority opinion. Each agent has two possible opinions, for instance, an ``up" opinion or \( +1 \) and a ``down" opinion or \( -1 \). To elaborate further, we outline the model algorithm as follows:

\begin{enumerate}
    \item The initial state of the system is prepared in a disordered state, where the number of agents with positive and negative opinions is equal.
    \item Microscopic interaction of agents within the model:
    \begin{enumerate}
        \item Model with independence: A group of agents is randomly selected from the population, and with a probability of $p$, each group member acts independently. Then, with the same probability of $1/2$, each member of the group changes its opinion to the opposite one.
        \item Model with anticonformity: A group of agents is randomly selected from the population, and with a probability of \( p \), all agents act in an anticonformist manner. If all agents within the group share the same opinion, they all change their opinions to the opposite one. \textcolor{black}{However, if the agents do not share the same opinion, they remain in their original state (no change occurs)}
    \end{enumerate}
    \item Alternatively, with a probability of $1 - p$, all agents choose to conform by following the majority opinion.
\end{enumerate}

We examine the model on several homogeneous networks, such as the complete graph, 2-D lattice, and 3-D lattice, and on several heterogeneous networks, such as the B-A, W-S, and E-R networks. All agent opinions are embedded in the network nodes, while the links or edges between nodes signify social connections. The complete graph depicts a network structure where every node is linked to every other node. In the complete graph, all agents are neighbors and can interact with each other with equal probability.

In each network, we select a group of agents with an odd size \( n \), and each group member interacts according to the algorithm described earlier. Specifically, in the complete graph, for each Monte Carlo step (time), a random group of \( n \) agents or nodes from anywhere in the network is selected, and they interact with each other based on the algorithm. For the complex networks, such as the B-A, W-S, and E-R networks and their combinations (multilayer networks), in each Monte Carlo step, one agent—let us call it the central agent or node—is randomly selected. Then, \( n-1 \) of its nearest neighbors are randomly selected. This \( n \)-member group interacts according to the model's algorithm. The group of \( n \) agents forms a `star-like formation,' where the central node acts as the `center,' and the neighboring agents act as `spokes' without necessarily being connected. 

\textcolor{black}{In the same manner as for complex networks, in the 2-D lattice, for each Monte Carlo step, a random agent, say \( S_{(i,j)} \), is selected, followed by the selection of its four nearest neighbors: \( S_{(i-1,j)}, S_{(i+1,j)}, S_{(i,j-1)} \), and \( S_{(i,j+1)} \). These five agents then interact according to the algorithm. In the 3-D lattice, for each Monte Carlo step, a random agent, say \( S_{(i,j,k)} \), is selected, followed by the selection of its six nearest neighbors: \( S_{(i-1,j,k)}, S_{(i+1,j,k)}, S_{(i,j-1,k)}, S_{(i,j+1,k)}, S_{(i,j,k-1)} \), and \( S_{(i,j,k+1)} \). These seven agents then interact according to the model's algorithm. In this paper, we consider all networks to be undirected and unweighted. Similar to the agent formation in the complex networks, the agent group of size \( n \) in these lattices forms a `star-like formation'.}

In a non-complete graph, each agent or node has the potential to influence up to its second-nearest neighbors via the central agent. Therefore, non-central agents may also compute the majority opinion. Such a scenario has also been applied to 2-D and 3-D lattices~\cite{chen2005majority}. From a social perspective, the central node serves as a bridge, facilitating indirect interactions between individuals or agents that are not directly connected. This structure promotes the spread of influence and consensus throughout the network by enabling interactions among second-nearest neighbors through an intermediary agent~\cite{centola2021change}.



For the model on the complete graph, we can conveniently perform analytical treatment to compute the order parameter (magnetization) of the model using the following formula:
\begin{equation}
    m = \dfrac{N_{+}-N_{-}}{N_{+}+N_{-}} = 2r - 1,
\end{equation}
where \( N_{+} \) and \( N_{-} \) represent the total number of agents with opinion \( +1 \) and \( -1 \), respectively, and \textcolor{black}{\( r = N_{+}/(N_{+} + N_{-})\)} denotes the fraction of opinion \(+1\), or the probability of finding an agent with \( +1 \) within the population. In the numerical simulation, we also compute the susceptibility \( \chi \) and Binder cumulant \( U \), which are defined as \cite{binder1981finite}:
\begin{align}
    \chi & = N \left[\langle m^2 \rangle - \langle m \rangle^2\right], \label{eq5}\\
    U & = 1 - \dfrac{1}{3}\dfrac{\langle m^4 \rangle}{\langle m^2 \rangle^2} \label{eq6}.
\end{align}
The parameters $m, \chi$, and $U$ are calculated once the simulation reaches an equilibrium state with a precision up to the order of \( 10^{-3} \).

We use finite-size scaling analysis to calculate the critical exponents corresponding to the order parameter $m$, susceptibility $\chi$, and Binder cumulant $U$. The finite-size scaling relations can be written as \cite{cardy1996scaling}:
\begin{align}
    p - p_{c} & = c\,N^{-1/\nu} \label{eq1},\\
    m & = \phi_{m}(x) N^{-\beta/\nu}, \\ 
    U & = \phi_{U}(x), \\
    \chi & = \phi_{\chi}(x) N^{\gamma/\nu} \label{eq2},
\end{align}
where \( \phi \) represents the dimensionless scaling function that fits the data near the critical point \( p_c \). The critical exponents \( \beta \), \( \gamma \), and \( \nu \) are important in the vicinity of the critical point \( p_c \). We can identify the critical point \( p_c \), where the system shifts between ordered and disordered phases, by pinpointing the intersection of the Binder cumulant curve \( U \) and the probability curve \( p \). This scaling analysis is important for understanding the universality class of the systems.

\section{ Result and Discussion}
\label{sec:result}
\subsection{Time evolution of the complete graph}
The evolution of the fraction \( r \) over time in the model can be obtained from the discrete-time master equation \cite{krapivsky2003dynamics}. For any given \( N \), the time evolution \( t \) is expressed as:
\begin{align} \label{eq:Eq8}
    r = r_0 + \dfrac{1}{N} \left[\rho_{+}(r)-\rho_{-}(r) \right],
\end{align}
where \( \rho^{+}(r) \) and \( \rho^{-}(r) \) represent the probabilities of the fraction \( r \) increasing and decreasing, respectively, at each time step \( \Delta t = 1/N \). For comparing the analytical and numerical simulations of the model on a complete graph, it is more appropriate to write the evolution of \( r \) for the limit \( N \gg 1 \), so that Eq.~\eqref{eq:Eq8} can be written as:
\begin{equation}\label{eq:time_ev_infty}
    \dfrac{\mathrm{d}r}{\mathrm{d}t} =  \rho^{+}(r)- \rho^{-}(r),
\end{equation}
which serves as the rate equation governing the fraction of opinion-up $r$. 

The form of \( \rho^{+}(r) \) and \( \rho^{-}(r) \) in Eq.~\eqref{eq:time_ev_infty} varies depending on the model. This study considers a scenario where three agents are randomly selected from the population to interact based on the outlined algorithm. We use the mean-field approach, assuming that the concentration of the global state is equal to the concentration of the local state, so the system's state can be represented by a single parameter—for example, the fraction of opinion \( r\).

For the model with independence, three selected agents ($n =3$) behave independently with a probability of $p$. All three agents change their opinion oppositely with a probability of $1/2$, transitioning from $-1$ to $+1$ if initially $-1$, and from $+1$ to $-1$ if initially $+1$. Hence, the probability density of the three agents changing their opinions from $+1$ to $-1$ is $3p/2 \left( N_{+}/N \right) $, and from $-1$ to $+1$ is $3p/2 \left( N_{-}/N \right)$. When the three agents do not exhibit independent behavior, with a probability of \(1-p\), they adopt the majority opinion. The opinion \( +1 \) increases when the three agents have a configuration of two opinions \( +1 \) and one opinion \( -1 \), such as \( ++- \), \( +-+ \), and \( -++ \). Similarly, the opinion \( +1 \) decreases when the three agents have a configuration of two opinions \( -1 \) and one opinion \( +1 \), such as \( --+ \), \( -+- \), and \( +-- \). Based on these configurations, the probability density of the three agents adopting the majority opinion \( +1 \) is:
\[
3(1-p)\frac{N_{+}}{N}\frac{N_{+}-1}{N-1}\frac{N_{-}}{N-2},
\]
and the probability density of adopting the majority opinion \( -1 \) is:
\[
3(1-p)\frac{N_{-}}{N}\frac{N_{-}-1}{N-1}\frac{N_{+}}{N-2}.
\]
Therefore, for any $N$, the total probabilities of the opinion $+1$ is increasing or decreasing, denoted by $\rho^{+}$ or $\rho^{-}$ respectively, can be written as follows:
  \begin{align}
    \rho^{+}& = 3\left[ \dfrac{p}{2} \dfrac{N_{-}}{N}+ \left(1 - p\right) \dfrac{N_{+}}{N} \dfrac{N_{+}-1}{N-1} \dfrac{N_{-}}{N-2}\right], \label{eq:rho_p}\\
     \rho^{-}&= 3 \left[ \dfrac{p}{2} \dfrac{N_{+}}{N}+ \left(1 - p\right) \dfrac{N_{-}}{N} \dfrac{N_{-}-1}{N-1} \dfrac{N_{+}}{N-2} \right]. \label{eq:rho_m}
\end{align}
And for $N\gg 1$, the equations \eqref{eq:rho_p} and \eqref{eq:rho_m} can be written as follows (see \ref{appendix1} for the general formulation):
\begin{align}
    \rho^{+}(r) & = 3 \left[ \dfrac{p}{2} \left(1 - r\right) + \left(1 - p\right) \left(1 - r\right) r^2\right], \label{eq10}\\
     \rho^{-}(r) & = 3 \left[ \dfrac{p}{2} r + \left(1 - p\right) r \left(1 - r\right)^2  \right
     ]. \label{eq11}  
\end{align}

For the model with anticonformity, three agents randomly selected will adopt anticonformist behavior with a probability of \( p \). These agents will change their opinions from \( -1 \) to \( +1 \) or vice versa when they share the same opinion. The opinion \( +1 \) will increase or decrease when all three agents have the same opinion \( -1 \) (i.e., \( --- \)) or \( +1 \) (i.e., \( +++ \)), with total probabilities of 
\[ 3p \cdot \frac{N_{-}}{N} \cdot \frac{N_{-}-1}{N-1} \cdot \frac{N_{-}-2}{N-2},  \] and 
\[3p \cdot \frac{N_{+}}{N} \cdot \frac{N_{+}-1}{N-1} \cdot \frac{N_{+}-2}{N-2}, \] respectively.

When the three agents do not adopt anticonformist behavior, they adopt the majority opinion with a probability of \( 1-p \), following the same probability configuration as in the independence model. Thus, the probability density of the opinion \( +1 \) increasing or decreasing for any \( N \) can be written as follows:
\begin{align}
    \rho^{+}(r) & = 3\left[p\,\dfrac{N_{-}}{N} \dfrac{N_{-}-1}{N-1} \dfrac{N_{-}-2}{N-2}+ \left(1-p \right) \dfrac{N_{+}}{N} \dfrac{N_{+}-1}{N-1} \dfrac{N_{-}}{N-2}\right], \label{eq:rho_p_an} \\
     \rho^{-}(r) & = 3\left[p\,\dfrac{N_{+}}{N} \dfrac{N_{+}-1}{N-1} \dfrac{N_{+}-2}{N-2}+ \left(1-p \right) \dfrac{N_{-}}{N} \dfrac{N_{-}-1}{N-1} \dfrac{N_{+}}{N-2} \right]. \label{eq:rho_m_an}  
\end{align}  
For $N\gg 1$, the equations \eqref{eq:rho_p_an} and \eqref{eq:rho_m_an} can be written as follows (see \ref{appendix2} for the general formulation):
\begin{align}
    \rho^{+}(r) & = 3 \left[ p\left(1-r\right)^{3}+ \left(1-p \right)  \left(1-r\right) r^{2}  \right], \label{eq12} \\
     \rho^{-}(r) & = 3 \left[ pr^{3}+\left(1-p \right) r \left(1-r\right)^{2}  \right]. \label{eq13}  
\end{align} 
Eqs.~\eqref{eq10}-\eqref{eq11} and \eqref{eq12}-\eqref{eq13} are essential for analyzing the system's state on the complete graph, especially in identifying order-disorder phase transitions. These equations allow us to understand the dynamics of opinion changes and the conditions under which the system transitions from an ordered to a disordered state at a certain critical point.

We can solve Eq.~\eqref{eq:time_ev_infty} to find an explicit expression for the fraction opinion-up $r$ at time $t$ for the models with independence and anticonformity. By substituting Eqs.~\eqref{eq10} and \eqref{eq11}, and integrating it, the fraction opinion $r$ for  the model with independence can be written as:
\begin{equation}\label{eq:time_indep_sol}
r(t, p, r_0) =\dfrac{1}{2} \left[ 1 \pm  \left(\dfrac{1-3p}{ 1-p+2e^{-3\left(1-3p\right) \left(t + A\right)}}\right)^{1/2} \right],
\end{equation}
where $A =  \ln[(1-2r_0)^2/(2(1-p)(r_0^2+r_0)+p]/[3(1-3p)]$ is a parameter that satisfies the initial condition of $r(t)$ at $t=0$.  Similarly, for the model with anticonformity, we obtain:
\begin{equation}\label{eq:time_anti_sol}
    r(t, p, r_0) = \dfrac{1}{2} \left[ 1 \pm  
 \left(\dfrac{1-4p}{1-4\,e^{-3\left(1-4p\right) \left(t+A \right)}}\right)^{1/2} \right],
\end{equation}
where $A = \ln\left[(1-2r_0)^2/(r_0^2-r_0+p)\right]/(1-4p)$. These equations provide analytic expressions for $r$ at time $t$ for both models, where $p$ represents the probability of agents adopting independence or anticonformity, and $r_0$ denotes the initial opinion fraction. For instance, when $p = 0$, both Eqs.~\eqref{eq:time_indep_sol} and \eqref{eq:time_anti_sol} converge to the same form, indicating the evolution of $r$ towards complete consensus states or completely disordered states, depending on the initial fraction $r_0$. Critical points occur at $p = 1/3$ for the model with independence and $p = 1/4$ for the model with anticonformity, where $r \to 1/2$, signifying complete disorder.

\begin{figure}[tb]
    \centering
    \includegraphics[width = \linewidth]{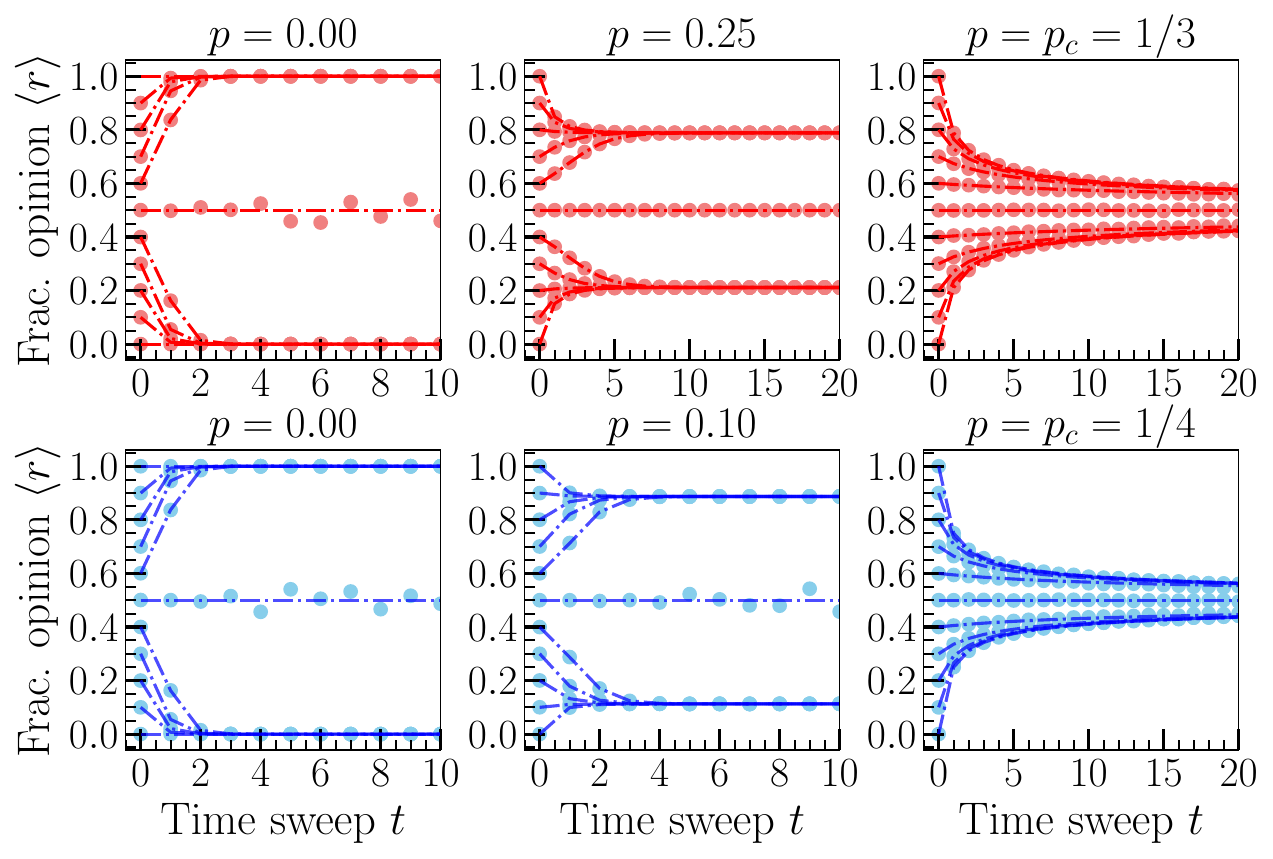}
    \caption{The comparison between analytical calculation (dashed lines) and numerical simulation (points) for both models with independence (red) and anticonformity (blue) across various probability values \( p \), based on Eqs.~\eqref{eq:time_indep_sol} and \eqref{eq:time_anti_sol}, respectively, is shown. At \( p = 0 \), all data points for \( r \) converge either to complete consensus with \( r = 1 \) (for \( r_0 > 1/2 \)) or \( r = 0 \) (for \( r_0 < 1/2 \)). For \( 0 < p < p_c \), all data points evolve towards two stable values \( r_{st} \), while at \( p = p_c \), all data converge to \( r \to 1/2 \) (representing a disordered state). The population size is \( N = 10^4 \), and each data point averages over 300 samples.}
    \label{fig:time_ev_indep}
\end{figure}

Figure \ref{fig:time_ev_indep} compares Eq.~\eqref{eq:time_indep_sol} (red) and Eq.~\eqref{eq:time_anti_sol} (blue) with numerical simulations for a large population (\( N = 10^4 \)) and different values of \( p \), demonstrating close alignment between analytical and numerical results. At \( p = 0 \) and \( |r_0| > 0.5 \), all initial fractions evolve towards complete consensus with \( r = 1 \) (all agents have the same opinion \( +1 \)) for \( r_0 > 0.5 \) and \( r = 0 \) (all agents have the same opinion \( -1 \)) for \( r_0 < 0.5 \). Additionally, at \( 0 < p < p_c \), \( r \) evolves towards two stable values, while at \( p = p_c \), \( r \) converges to \( 1/2 \), representing complete disorder.

\subsection{Phase diagram and critical exponents of the model on the complete graph}

To investigate the order-disorder phase transition of the model, we consider the stationary condition of Eq.~\eqref{eq:time_ev_infty}, where \(\mathrm{d}r/\mathrm{d}t = 0\) or \(\rho^{+} = \rho^{-}\). For the model with independence, this yields three stationary solutions: \( r_{1} = 1/2 \) and \( r_{2,3} = \dfrac{1}{2} \left( 1 \pm \left(\dfrac{1-3p}{1-p}\right)^{1/2} \right) \). Consequently, the order parameter \( m \) is:
\begin{align}\label{eq:phase_indep}
m_{2,3} & = \pm \left(\dfrac{1-3p}{1-p}\right)^{1/2}.
\end{align}
The critical point occurs at \( p_c = 1/3 \), where \( m_{2,3} = 0 \). Similarly, for the anticonformity model, the stationary condition for the opinion fraction \(\mathrm{d}r/\mathrm{d}t = 0\) yields three stationary states: \( r_1 = 1/2 \) and \( r_{2,3} = \dfrac{1}{2} \left(1 \pm \left(1 - 4p\right)^{1/2}\right) \). Consequently, the order parameter \( m \) is:
\begin{align}\label{eq:phase_anti}
 m_{2,3} = \pm \left(1-4p\right)^{1/2},
\end{align}
and hence, the critical point for the anticonformity model occurs at \( p_c = 1/4 \). Both Eqs.~\eqref{eq:phase_indep} and \eqref{eq:phase_anti} can be expressed as power law in terms of \( p \), where \( m \sim (p - p_c)^\beta \), with \(\beta = 1/2\), typical of the critical exponent of the mean-field Ising model \cite{stanley1971phase}.

\begin{figure}[tb]
    \centering
    \includegraphics[width = \linewidth]{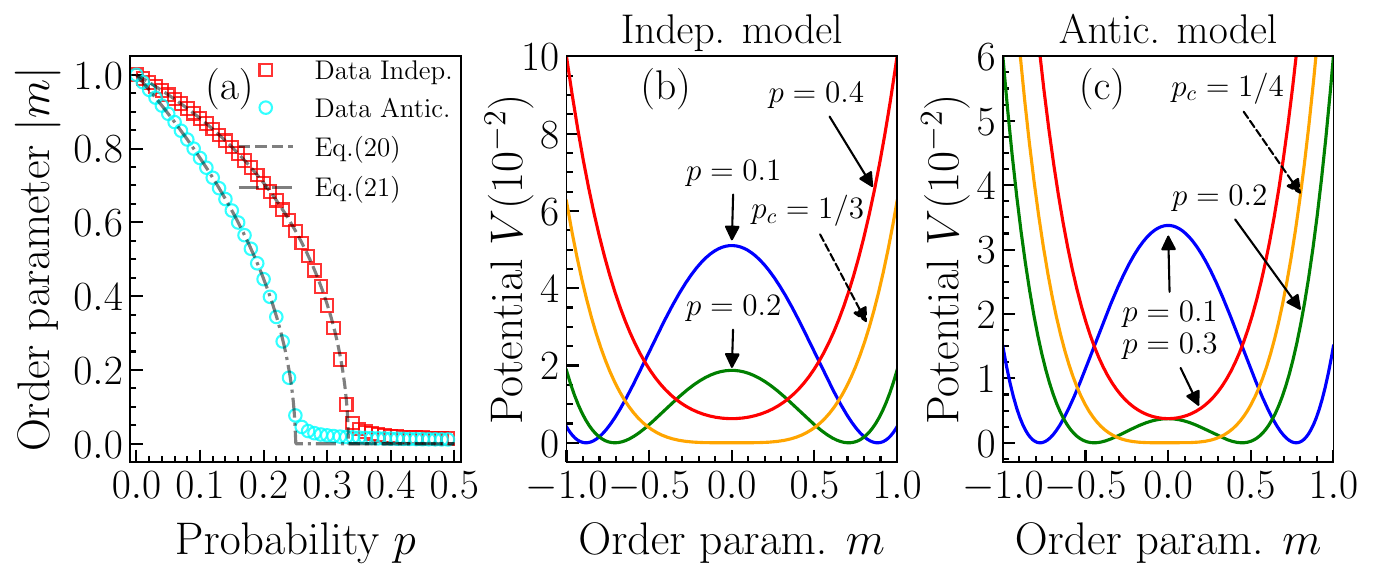}
    \caption{ The phase diagram [panel (a)] of the model on the complete graph for models with independence and anticonformity is shown. Analytical results from Eqs.~\eqref{eq:phase_indep} and \eqref{eq:phase_anti} are compared with the Monte Carlo simulation data and agree very well. Both models undergo continuous phase transitions, revealing critical points at \( p_c = 1/3 \) for the model with independence and \( p_c = 1/4 \) for the model with anticonformity. Panels (b) and (c) illustrate the effective potential described by Eqs.~\eqref{eq:pot_inde} and \eqref{eq:pot_anti}, demonstrating bistable for $p < p_c$ and monostable behaviors for $p > p_c$, indicating the continuous phase transition is occurred.}
    \label{fig:phase_dia}
\end{figure}

As previously discussed, the topology of the complete graph can be approximated analytically using a mean-field approach. To validate these analytical results, Monte Carlo simulations were performed with a population size of $N = 10^4$. The results, illustrated in Fig.~\ref{fig:phase_dia} [panel (a)], demonstrate a close match between the analytical calculations and the Monte Carlo simulations. These data indicate that the model undergoes a continuous phase transition with a critical point \( p_c = 1/3 \) for the independence model and \( p_c = 1/4 \) for the anticonformity model.

Another method to analyze the order-disorder phase transition is the effective potential, obtained through integration of the effective force. Traditionally, the effective potential is derived from the effective force using the formula $V(r)_{\text{eff}} = -\int f(r)_{\text{eff}} \, \mathrm{d}r$. Here, $f(r)_{\text{eff}} = \rho^{+} (r) - \rho^{-}(r)$ represents the force that drives opinion change during the dynamics process. For the independence model on the complete graph, the effective potential (in terms of $m$) can be written as follows:
\begin{align} \label{eq:pot_inde}
V_{\text{indep.}} =\frac{3}{32 \left(1-p\right)} \left(1-3p-\left(1-p\right) m^{2}\right)^{2}.
\end{align}
And for the model with anticonformity, the effective potential can be written as follows:
\begin{align}\label{eq:pot_anti}
V_{\text{antic.}} = \frac{3}{32}\left(1-4p-m^{2}\right)^{2}.
\end{align}
Plots of Eqs.~\eqref{eq:pot_inde} and \eqref{eq:pot_anti} are shown in panels (b) and (c) of Fig.~\ref{fig:phase_dia}. For both potentials, there are bistable states for $p < p_c$, a bistable-monostable transition at $p = p_c$, revealing the model's critical point, and the system is monostable for $p >p_c$, indicating a continuous phase transition at $p_c$.

The critical point of the model can also be analyzed using Landau's theory. According to this theory, the potential can be expanded in terms of the magnetization $m$ as $V = \sum_i V_i m^i$, where $V_i$ generally depends on thermodynamic parameters \cite{landau1937theory}. In this model, $V_i$ can be influenced by noise parameters, such as the probabilities of independence and anticonformity. The Landau potential $V$ exhibits symmetry under the inversion of the order parameter, $m \to -m$. Consequently, only the even terms of the potential are considered. Therefore, the Landau potential takes the form:
\begin{equation}\label{eq:land_pot}
V = V_2 m^2 + V_4 m^4 +  \cdots
\end{equation}

Understanding the terms $V_2$ and $V_4$ is sufficient for analyzing the model's phase transition using the potential $V$. The critical point can be determined by setting $V_2 = 0$, and the nature of the phase transition is characterized by $V_4 (p_c)$, where $V_4 (p_c) \geq 0$ denotes a continuous phase transition, while $V_4 (p_c) < 0$ indicates a discontinuous phase transition. By comparing Eq.~\eqref{eq:land_pot} with Eqs.~\eqref{eq:pot_inde} and \eqref{eq:pot_anti}, we can determine $V_2$ and $V_4$ for both the independence and anticonformity models. For the independence model, we obtain $V_2(p) = 3 \left(1-3p \right)/8 $ and $V_4(p) = 9\left(1-p\right)/4 $. Meanwhile, for the anticonformity model, we obtain $V_2(p) = -3\left(1-4p\right)/8$ and $ V_4(p) = 9/4$. Hence, the critical points $p_c$ align with those obtained from the previous analysis: $p_c = 1/3$ for the independence model and $p_c = 1/4$ for the anticonformity model. Furthermore, $V_4 (p_c) \geq 0$ for both models confirms their continuous phase transition. 

\begin{figure}[tb]
    \centering
    \includegraphics[width = \linewidth]{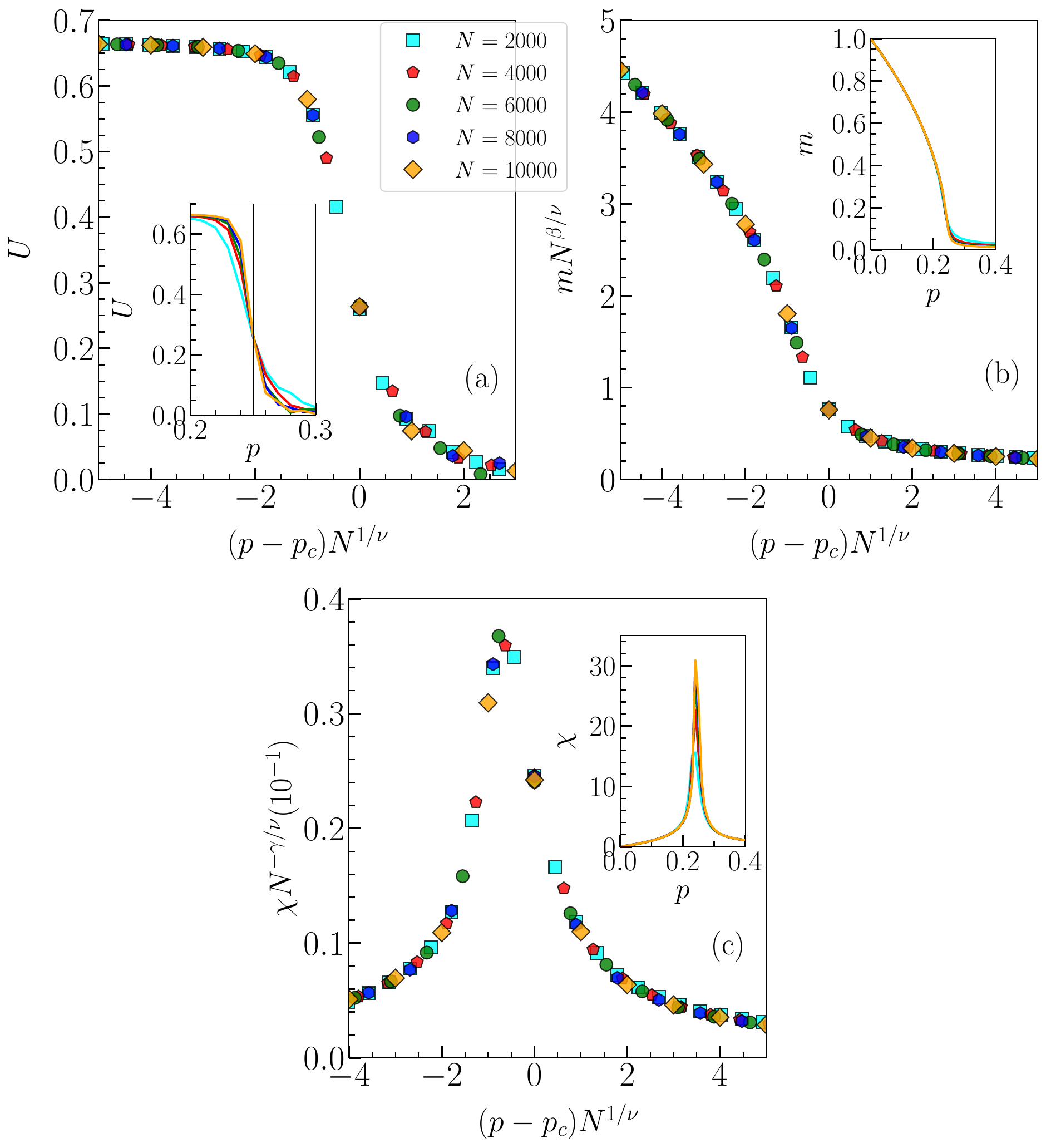}
    \caption{The M-C simulation results of the model on the complete graph for the order parameter \( m \), susceptibility \( \chi \), and Binder cumulant \( U \) across various population sizes \( N \). The critical point is identified from the crossing lines of the Binder cumulant \( U \) versus probability \( p \) which is found at \( p_c \approx 0.250 \) [inset graph of the panel (a)], validating the analytical result in Eq.~\eqref{eq:phase_anti}. The best critical exponents facilitating the collapse of all data near the critical point \( p_c \) are \( \beta \approx 0.5 \), \( \nu \approx 2.0 \), and \( \gamma \approx 1.0 \). These values suggest that the model belongs to the mean-field Ising model class.}
    \label{fig:crit_cg}
\end{figure}

The model's critical points and critical exponents can be estimated numerically using finite-size scaling relations [Eqs.~\eqref{eq1}-\eqref{eq2}]. By varying the population size $N$ from $2 \times 10^3$ to $10^4$, we compute the magnetization $m$, susceptibility $\chi$, and Binder cumulant $U$ as shown in Fig.~\ref{fig:crit_cg}. Each data point is averaged over $10^5$ independent realizations to ensure accurate results. In Fig.~\ref{fig:crit_cg}, the inset graphs display standard plots, while the main graphs present the scaling plots of the model. The critical point is determined using the Binder method by observing the crossing of lines between the Binder cumulant $U$ and the probability of anticonformity $p$. In this instance, the critical point is estimated to be $p_c \approx 0.251$ [inset graph of panel (a)], consistent with the analytical result in Eq.~\eqref{eq:phase_anti}, namely $p_c = 1/4$.

The plots in Fig.~\ref{fig:crit_cg} show the dynamics of the scaled parameters for the model with anticonformity. The best critical exponents obtained from fitting the data for various values of $N$ are $\beta \approx 0.5$, $\nu \approx 2.0$, and $\gamma \approx 1.0$. It is important to note that although $\beta = 1/2$ and $\gamma = 1$ are the same with the usual critical exponents for the mean-field Ising model, $\nu = 2.0$ does not fit in this pattern. However, a direct connection exists between $\nu$ and the critical dimension $d_c = 4$ of the mean-field Ising model, expressed as $\nu = d_c \nu' = 2$, where $\nu' = 1/2$ is an effective exponent. This result is also observed in several discrete dynamic models \cite{crokidakis2015inflexibility, muslim2020phase, muslim2022opinion, muslim2022phase, oestereich2023phase, biswas2012disorder}. These critical exponents suggest that the model belongs to the mean-field Ising universality class. Notably, identical critical exponents are obtained for the independence model, indicating a similarity between the independence and anticonformity models. Furthermore, these models resemble well-known models like the Sznajd \cite{muslim2022opinion} and kinetics exchange models~\cite{crokidakis2014phase,biswas2023social}. This finite-size scaling analysis provides robust numerical evidence supporting the model's critical point and exponents, validating the analytical findings and classifying the model within the mean-field Ising universality class.

\subsection{Critical exponents of the model on the 2-D lattice}
\begin{figure}[tb]
    \centering
    \includegraphics[width = \linewidth]{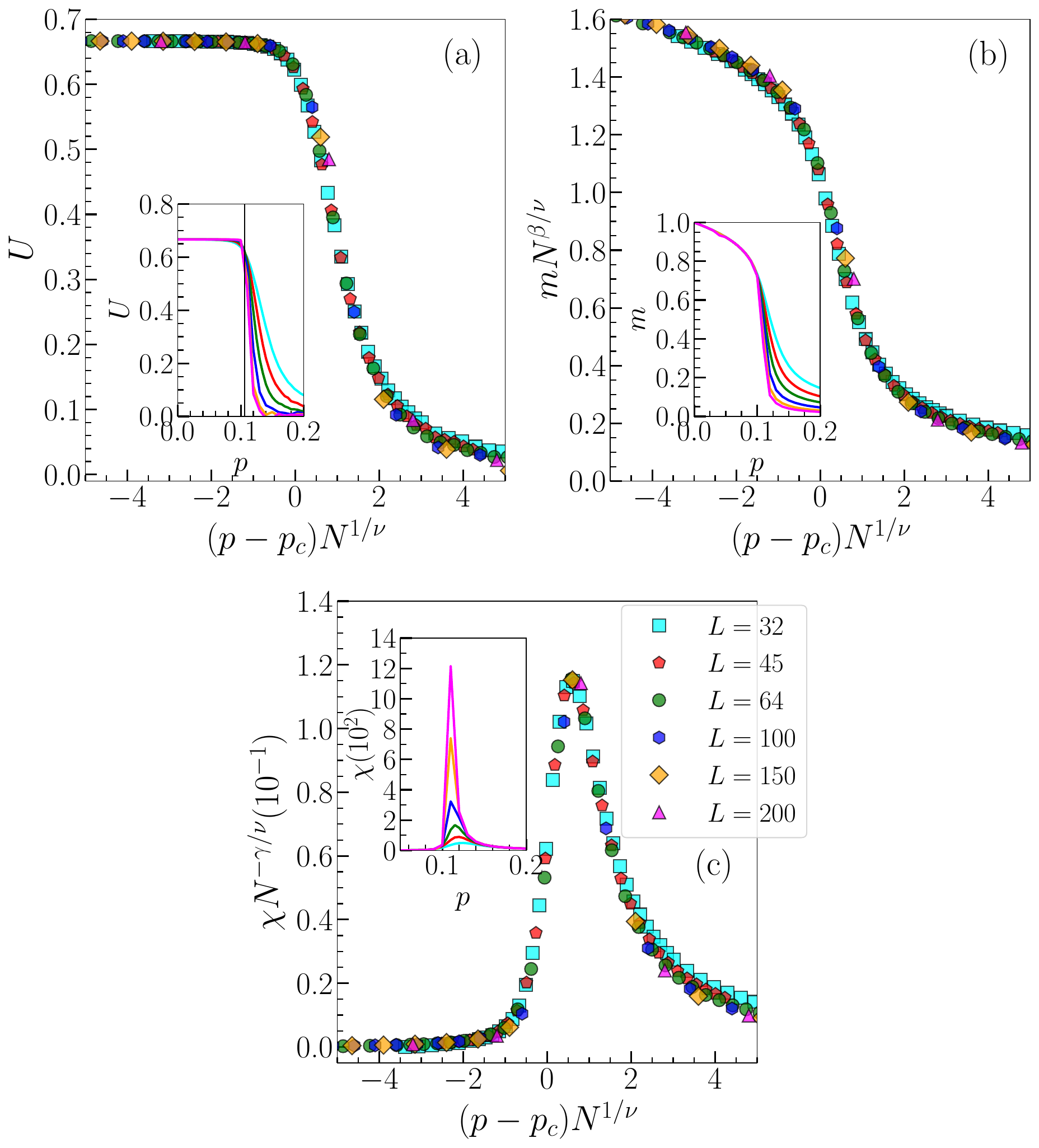}
    \caption{The M-C simulation results of the model on the 2-D lattice for the model with independence for the order parameter \( m \), susceptibility \( \chi \), and Binder cumulant \( U \) across various population sizes \( N = L^2\). The model undergoes a continuous phase transition with a critical point at \( p_c \approx 0.106 \). The best critical exponents of the model are \( \beta \approx 0.125 \), \( \nu \approx 1.0 \), and \( \gamma \approx 1.75 \). These results suggest that the model falls into the same universality class as the 2-D Ising model.}\label{fig:phase_tranisiton_indep}
\end{figure}

We explored various population sizes denoted as $N = L^2$, where $L$ takes values of $32, 45, 64, 100, 150$, and $200$, to investigate the model's critical point and critical exponents on the 2-D lattice. The numerical results concerning the order parameter $m$, susceptibility $\chi$, and Binder cumulant $U$ are depicted in Fig.~\ref{fig:phase_tranisiton_indep}. The critical point, marking the instance of a continuous phase transition in the model, is identified at $p_c \approx 0.106$ [as observed in the inset panel (a) of Fig.~\ref{fig:phase_tranisiton_indep}]. By employing finite-size scaling relations described in Eqs.~\eqref{eq1}-\eqref{eq2}, we determined the critical exponents that give the best description of the data. These critical exponents are $\beta \approx 0.125, \gamma \approx 1.75,$ and $\nu \approx 1.0$. \textcolor{black}{These values suggest similarities with the 2-D majority vote model with a single opinion change \cite{de1992isotropic}, as well as with the 2-D Sznajd model \cite{AZHARI2024130042, calvelli2019phase}, and align with the universality class of the 2-D Ising model \cite{stanley1971phase}.}

\begin{figure}[tb]
    \centering
    \includegraphics[width = \linewidth]{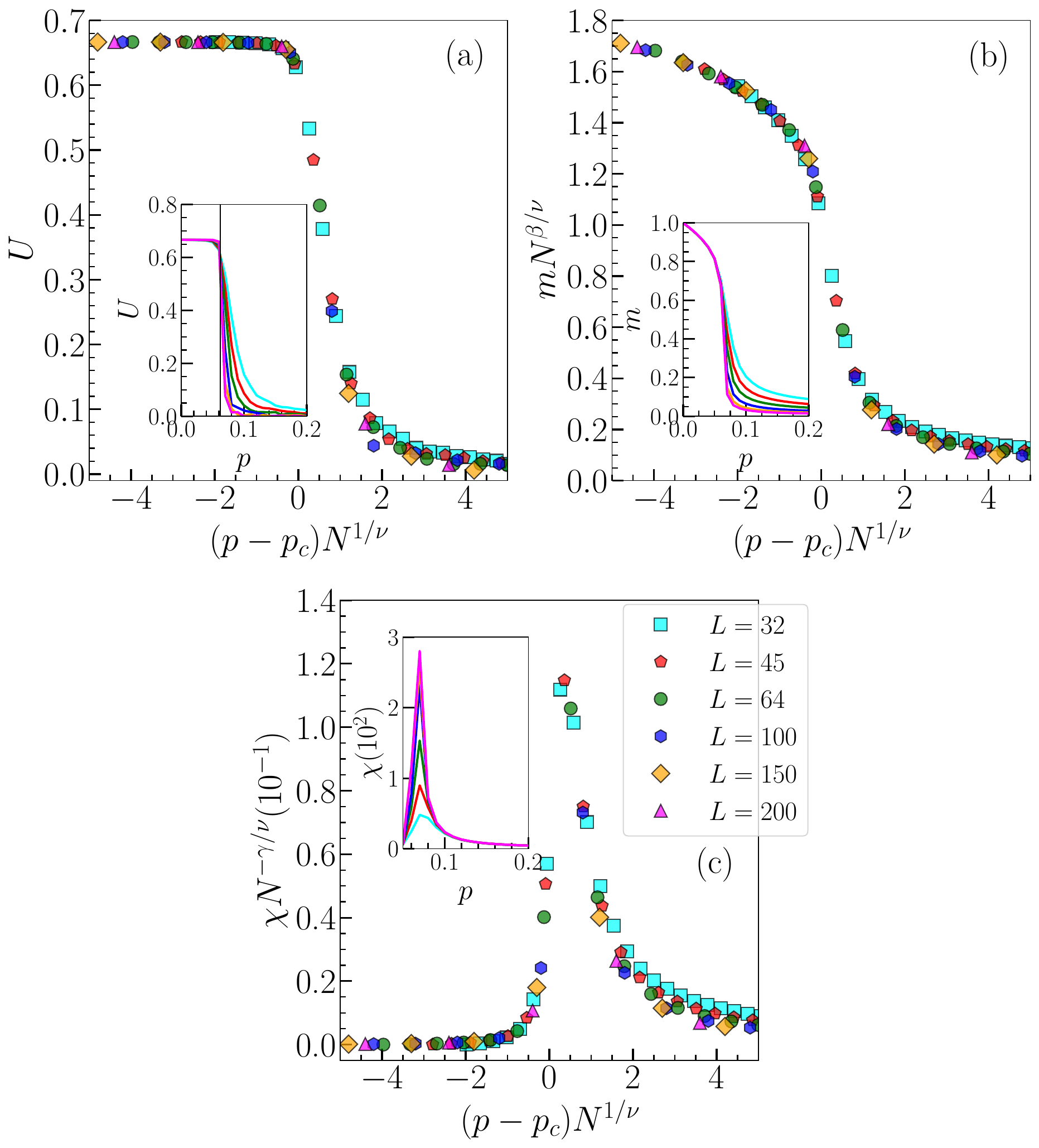}
    \caption{The M-C simulation results of the model on the 2-D lattice for the model with anticonformity for the order parameter \( m \), susceptibility \( \chi \), and Binder cumulant \( U \) across various population sizes \( N = L^2\). The model undergoes a continuous phase transition with a critical point at \( p_c \approx 0.062 \). The best critical exponents of the model are \( \beta \approx 0.125 \), \( \nu \approx 1.0 \), and \( \gamma \approx 1.75 \). These results suggest that the model falls into the same universality class as the 2-D Ising model.}\label{fig:phase_tranisiton_anti}
\end{figure}

We extended our analysis to incorporate the model with anticonformity and found that it also undergoes a continuous phase transition. The critical point for the model with anticonformity is \( p_c \approx 0.062 \), as shown in the inset panel (a) of Fig.~\ref{fig:phase_tranisiton_anti}. Notably, our investigations yielded similar critical exponents for this model: $\beta \approx 0.125, \gamma \approx 1.75,$ and $\nu \approx 1.00$. These shared critical exponents suggest that both the model with independence and the model with anticonformity exhibit analogous behavior, indicating their belonging to the same universality class. Our results align these models with the two-dimensional Ising universality class. The critical exponents satisfy the hyperscaling relation \(\nu d = 2 \beta + \gamma\), where \(d = 2\), the spatial dimension of the model.

\subsection{Critical exponents  of the model on the 3-D lattice}

\begin{figure}[tb]
    \centering
    \includegraphics[width = \linewidth]{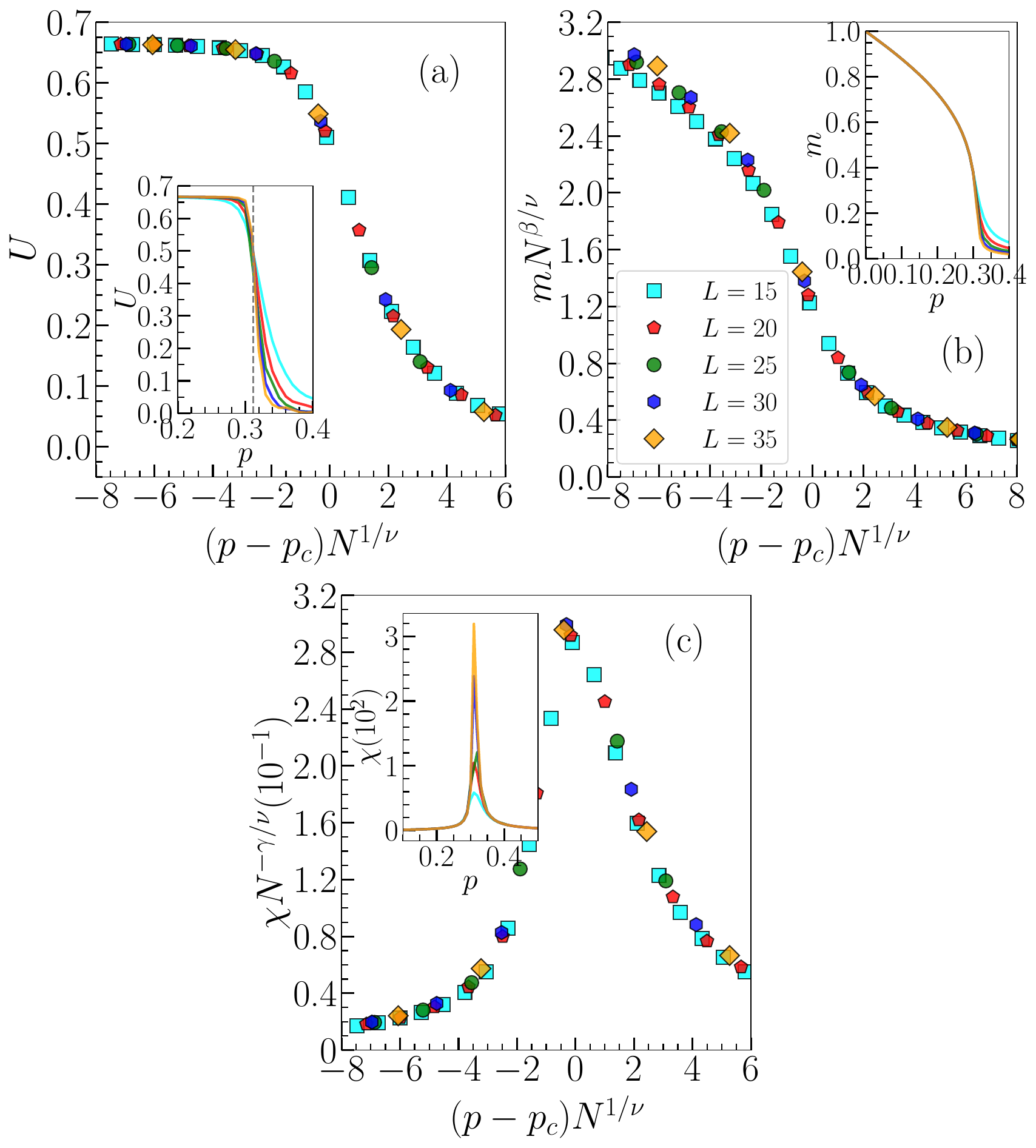}
    \caption{The M-C simulation results of the model on the 3-D lattice for the model with independence for the order parameter \( m \), susceptibility \( \chi \), and Binder cumulant \( U \) across various population sizes \( N = L^3\). The model undergoes a continuous phase transition with a critical point at \( p_c \approx 0.311 \). The best critical exponents of the model are \( \beta \approx 0.326 \), \( \nu \approx 0.630 \), and \( \gamma \approx 1.237 \) . These results suggest that the model falls into the same universality class as the 3-D Ising model. }\label{fig:3D_indep}
\end{figure}

We investigated the model using different population sizes \( N = L^3 \), where the linear dimensions \( L \) varied from 15 to 35. Similar to the 2-D lattice model, we used periodic boundary conditions. The numerical results for the Binder cumulant \( U \), order parameter \( m \), and the susceptibility \( \chi \) of the model with independence are shown in Fig.~\ref{fig:3D_indep}. Each data point averages over \( 10^6 \) independent realizations. Our results indicate that the model undergoes a continuous phase transition, with the critical point occurring at \( p_c \approx 0.311 \). Using finite-size scaling analysis near the critical point \( p_c \), we determined the critical exponents of the model to be \( \nu \approx 0.630 \), \( \beta \approx 0.326 \), and \( \gamma \approx 1.237 \). These critical exponent values are consistent with those of the 3-D Ising model, suggesting that this model belongs to the same universality class as the 3-D Ising model \cite{stanley1971phase}.

\begin{figure}[tb]
    \centering
    \includegraphics[width = \linewidth]{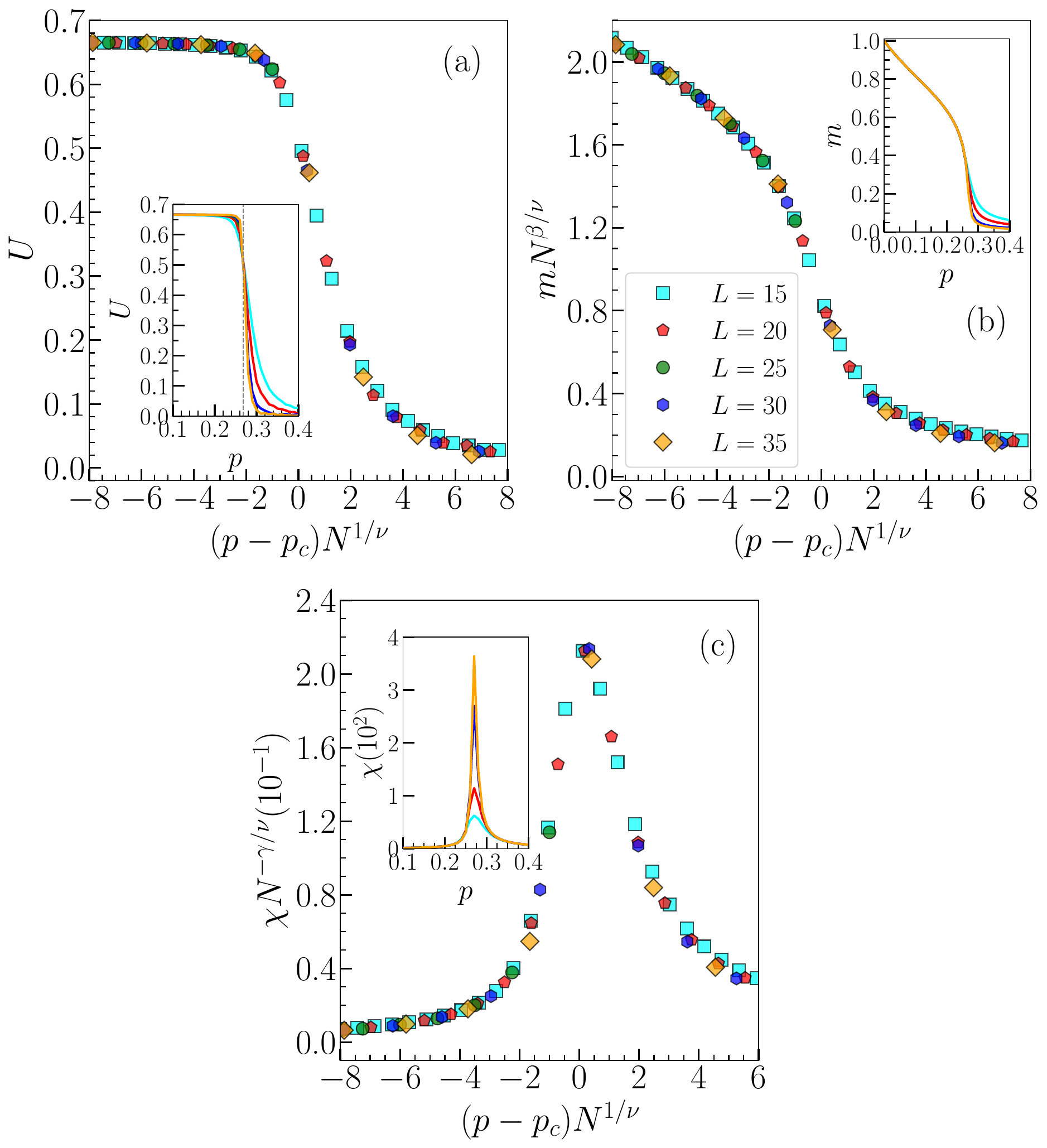}
    \caption{The M-C simulation results of the model on the 3-D lattice for the model with anticonformity for the order parameter \( m \), susceptibility \( \chi \), and Binder cumulant \( U \) across various population sizes \( N = L^3\). The model undergoes a continuous phase transition with a critical point at \(  p_c \approx 0.268 \). The best critical exponents of the model are \( \beta \approx 0.326 \), \( \nu \approx 0.630 \), and \( \gamma \approx 1.237 \) . These results suggest that the model falls into the same universality class as the 3-D Ising model.}\label{fig:3D_anti}
\end{figure}

Our analysis of the model with anticonformity reveals a second-order phase transition with a critical point at \( p_c \approx 0.268 \), as shown in Fig.~\ref{fig:3D_anti}. In this model, we obtained similar critical exponents to those of the model with independence, yielding values of \( \nu \approx 0.630 \), \( \beta \approx 0.326 \), and \( \gamma \approx 1.237 \). These results suggest that both the models with independence and anticonformity are identical and belong to the same universality class. Notably, these critical exponents remain consistent across various datasets for different system sizes \( N \), indicating their universality. It is important to note that the critical exponents of both models satisfy the hyperscaling relation \( \nu d = 2\beta + \gamma \), where $d = 3$, the spatial dimension of the model.

\subsection{Critical exponents of the model on the heterogeneous networks}
Compared to the homogeneous networks mentioned earlier, heterogeneous networks such as Watts-Strogatz (W-S), Albert-Barabási (A-B), and Erdős-Rényi (E-R) networks better reflect real social networks~\cite{albert2002statistical, newman2018networks}. These three types of networks have been extensively studied across various research areas and applied to understand diverse social phenomena, such as epidemic processes~\cite{pastor2015epidemic}, the analysis of the structure and characteristics of scientific collaboration networks~\cite{newman2001structure}, including in fields like medicine~\cite{barabasi2023neuroscience}.

In the heterogeneous networks, we examine the B-A network with an average node degree of \( \langle k \rangle = 6 \), which corresponds to a degree distribution exponent \( 2 < \gamma < 3 \). For the W-S network, we analyze a network with an average node degree of \( \langle k \rangle = 4 \) and a small rewiring probability \( \alpha = 0.1 \), which tends to depict sparse networks. For the E-R network, we consider a small connection probability \( p = 0.002 \) or an average node degree of \( \langle k \rangle \sim Np= 20 \), where the agent or node population \( N = 10^4 \). The selection of these network properties ensures that each agent has at least two nearest neighbors (i.e., directly connected neighbors) while maintaining a relatively small difference in \( \langle k \rangle \) across the three networks. This configuration allows for a meaningful interaction structure in the model, ensuring that agents are not isolated and can effectively participate in the dynamics of opinion changes within the network.

\begin{figure}[tb]
    \centering
    \includegraphics[width = \linewidth]{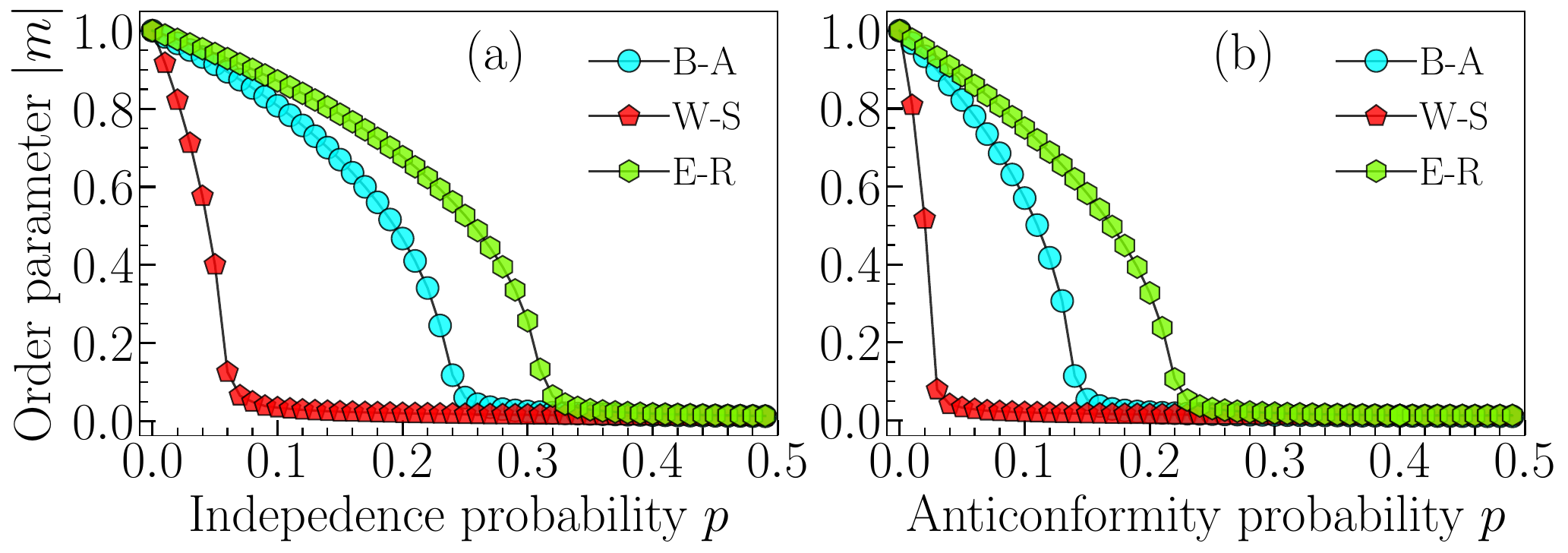}    \caption{The Monte Carlo simulation results for the models with independence [panel (a)] and anticonformity [panel (b)] on the Barabási-Albert (B-A) network with \(\langle k \rangle = 6\), the Watts-Strogatz (W-S) network with rewiring probability \(\alpha = 0.1\) and \(\langle k \rangle = 4\), and the Erdős-Rényi (E-R) network with connection probability \(p = 0.002\) or $\langle k \rangle \sim 20 $ show that the model undergoes continuous phase transitions. The population size is \(N = 10^4\), and each data point represents the average of \(10^5\) independent realizations.}\label{fig:complex_graph}
\end{figure}

In all three networks, three randomly chosen agents ($n=3$) interacted according to the model's algorithm. The population size was set to \( N = 10^4 \), and each data point represents the average of \( 10^5 \) independent realizations. Our numerical results for the order parameter \( m \) are shown in Fig.~\ref{fig:complex_graph}.  One can see that the model undergoes a continuous phase transition in all three networks, each with different critical points. At \( p = 0 \), the system exhibits complete order with \( |m|=1 \) (complete consensus with all members having the same opinion). The value of \(|m|\) decreases as \(p\) increases and approaches zero near the critical point \(p_c\). For \(p < p_c\), the system is in a state of consensus with a majority-minority opinion existing, and for \(p > p_c\), the system is in a state of polarization. Interestingly, for all three networks, the critical point for the model with anticonformity is smaller compared to the model with independence on the same network, consistent with the results on homogeneous networks.

\textcolor{black}{For the model on the W-S network, particularly for the model with anticonformity, it can be observed in panel (b) of Fig. \ref{fig:complex_graph} that the order-disorder transition occurs at a small value of \( p \). This phenomenon is due to the low rewiring probability of the network, \( \alpha = 0.1 \), which makes the network tend to be more homogeneous or regular. When \( \alpha = 0.0 \), the W-S network behaves like a 1-D lattice, where no phase transition is observed. The phase diagram of the model, especially the anticonformity model on the W-S network for different values of \( \alpha \), is explored further in~\ref{app:b.1}.}

\begin{figure}[tb]
    \centering
    \includegraphics[width = \linewidth]{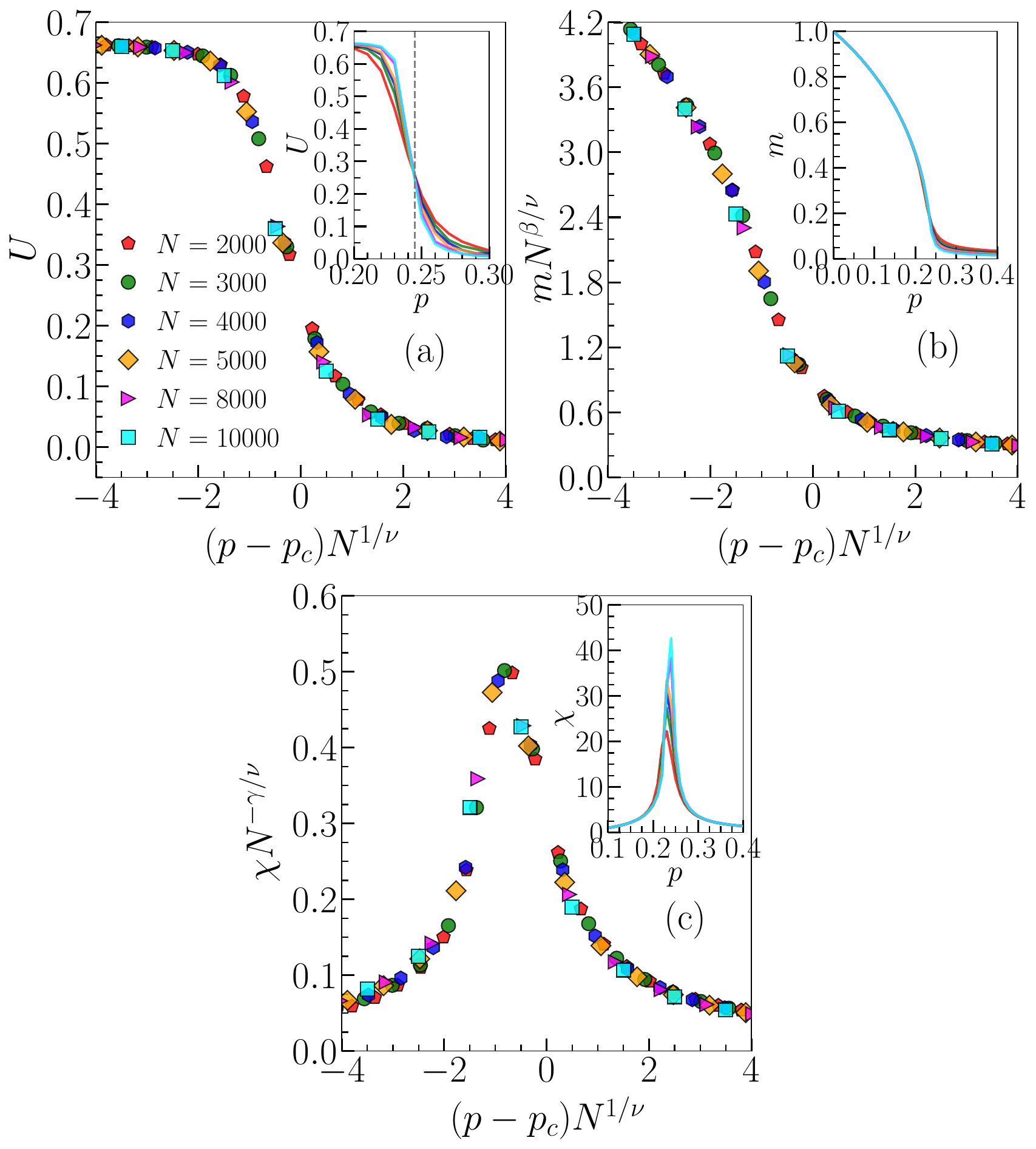}
    \caption{The M-C simulation results of the model on the B-A network for the model with independence for the order parameter \( m \), susceptibility \( \chi \), and Binder cumulant \( U \) across various population sizes \( N \). The model undergoes a continuous phase transition with a critical point at \(  p_c \approx 0.245 \). The best critical exponents of the model are \( \beta \approx 0.5 \), \( \nu \approx 2.0 \), and \( \gamma \approx 1.0 \) . These results suggest that the model is identical to the model on the complete graph and falls into the mean-field Ising universality class.}\label{fig:Fig09}
\end{figure}

We analyzed the critical exponents of the model using finite-size scaling in Eqs.~\eqref{eq1}-\eqref{eq2}. The numerical results for the model with independence on the B-A network are shown in Fig.~\ref{fig:Fig09}. The model exhibits a continuous phase transition with a critical point at \( p_c \approx 0.245 \) [see the inset graph in panel (a)]. The best fit for the critical exponents across different \( N \) values are \( \nu \approx 2.0 \), \( \gamma \approx 1.0 \), and \( \beta \approx 0.5 \). These results suggest that the model \textcolor{black}{belongs to the same universality class as the majority vote model on scale-free and E-R networks with a single opinion change~\cite{kim2021majority}, and also shares the universality class of the model on a complete graph, aligning with the mean-field Ising model.}

\begin{figure}[tb]
    \centering
    \includegraphics[width = 0.5\linewidth]{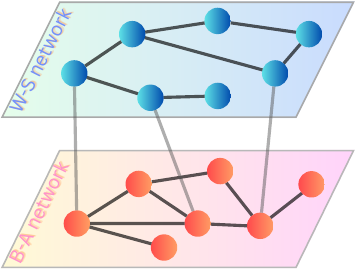}
    \caption{The combination of the B-A network and the W-S network, where several nodes between the two networks are interconnected, denoted by \( N_c \), represents a topology that can model two communities or groups, each carrying its characteristics. In this setup, some individuals in each group can communicate with individuals in the other group. \textcolor{black}{In Figs.~\ref{fig:combined_inde} and \ref{fig:combined_anti}, we use the properties of the B-A and W-S networks, with the B-A network having an average degree of connection \( \langle k \rangle = 6 \) and the W-S network having a rewiring probability \( \alpha = 0.1 \) and an average degree of connection \( \langle k \rangle = 4 \). The number of interconnected agents, or nodes \( N_c \), in each network is fixed at $50$ for all network sizes $N$.}}\label{fig:combined_network}
\end{figure}

\textcolor{black}{We also analyzed the model on a combination of the B-A and W-S networks (a multilayer network), where several nodes between the two networks are interconnected, as shown in Fig.~\ref{fig:combined_network}. This multilayer network was constructed as follows: a B-A network with a preferential attachment mechanism and a W-S network based on a ring lattice with a rewiring probability \( \alpha \) were generated. After both networks were formed, a subset of nodes (denoted as \( N_c \)) from each network was connected to nodes in the other network. No explicit probability was defined for the likelihood of inter-network connections; instead, we specified the number of nodes in each network to be connected to the other network. The combination of these two networks resulted in a multilayer network that retains the characteristics of both the B-A and W-S networks, with some nodes being interconnected between the two. These interconnections allow for the diffusion of information across both networks. This multilayer network can also represent two communities or groups where each member can communicate with others. Within this network, we also examined interactions among three agents, who interact with each other according to the model’s algorithm. We ensured that each node in the combined network had at least two directly connected neighbors.}

\begin{figure}[tb]
    \centering
    \includegraphics[width = \linewidth]{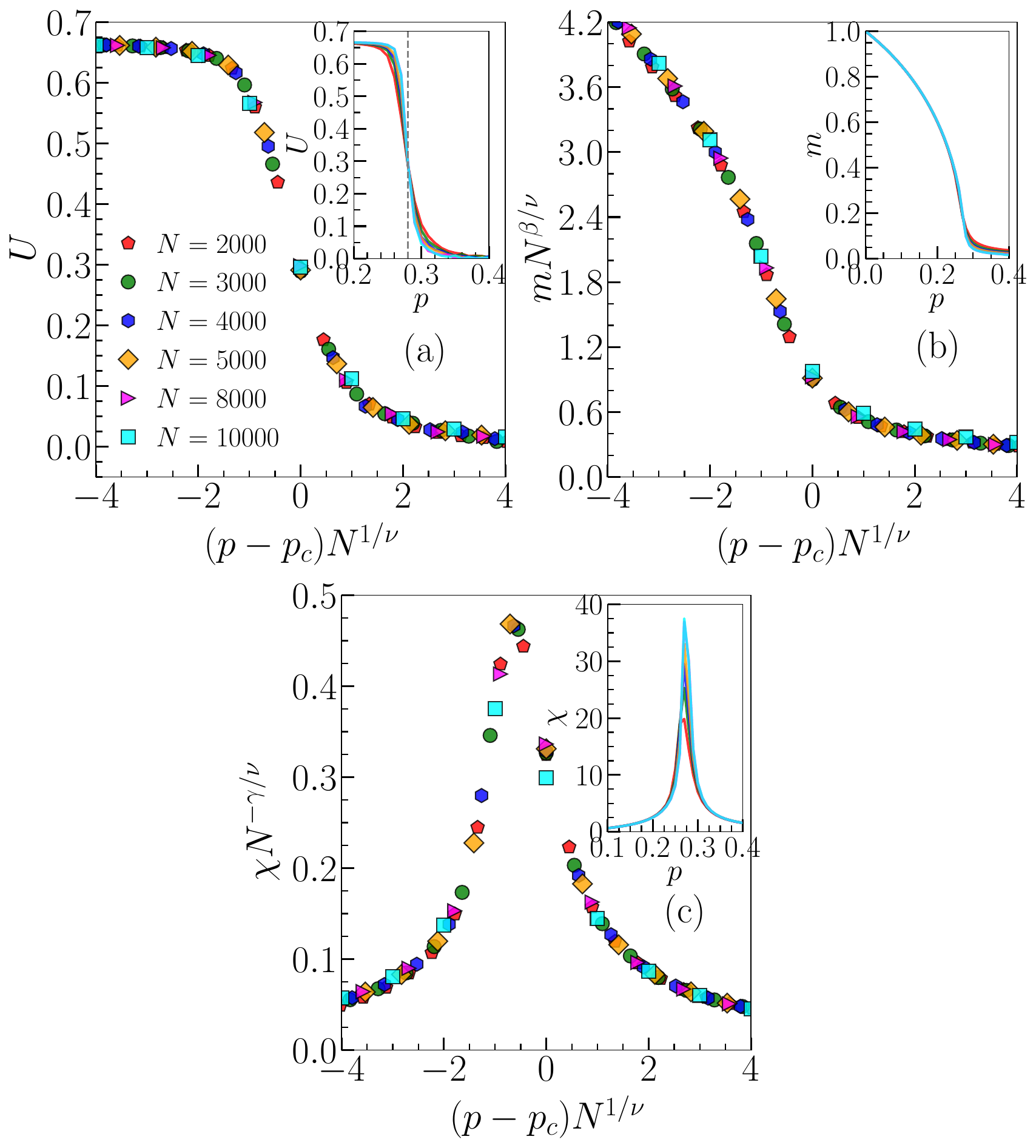}
    \caption{The M-C simulation results of the model on the combined B-A and W-S network for the model with independence for the order parameter \( m \), susceptibility \( \chi \), and Binder cumulant \( U \) across various population sizes \( N \). The model undergoes a continuous phase transition with a critical point at \(  p_c \approx 0.280 \). The best critical exponents of the model are \( \beta \approx 0.5 \), \( \nu \approx 2.0 \), and \( \gamma \approx 1.0 \) . These results suggest that the model is identical to the model on the complete graph and falls into the mean-field Ising universality class.}\label{fig:combined_inde}
\end{figure}

The numerical results for \( m \), \( \chi \), and \( U \) for the model with independence on this combined network are shown in Fig.~\ref{fig:combined_inde}. It can be seen that the model undergoes a continuous phase transition with a critical point at \( p_c \approx 0.280 \). Interestingly, the finite-size scaling analysis results show that the model has critical exponents of \( \nu \approx 2.0 \), \( \gamma \approx 1.0 \), and \( \beta \approx 0.5 \) across different \( N \). This indicates that although the topology of the combined network is constructed from two different heterogeneous networks, the critical phenomena and scale invariance are identical to the model on a complete graph, the B-A network, and the mean-field Ising model. Therefore, it can be said that this model belongs to the same universality class as the mean-field Ising model.

\begin{figure}[tb]
    \centering
    \includegraphics[width = \linewidth]{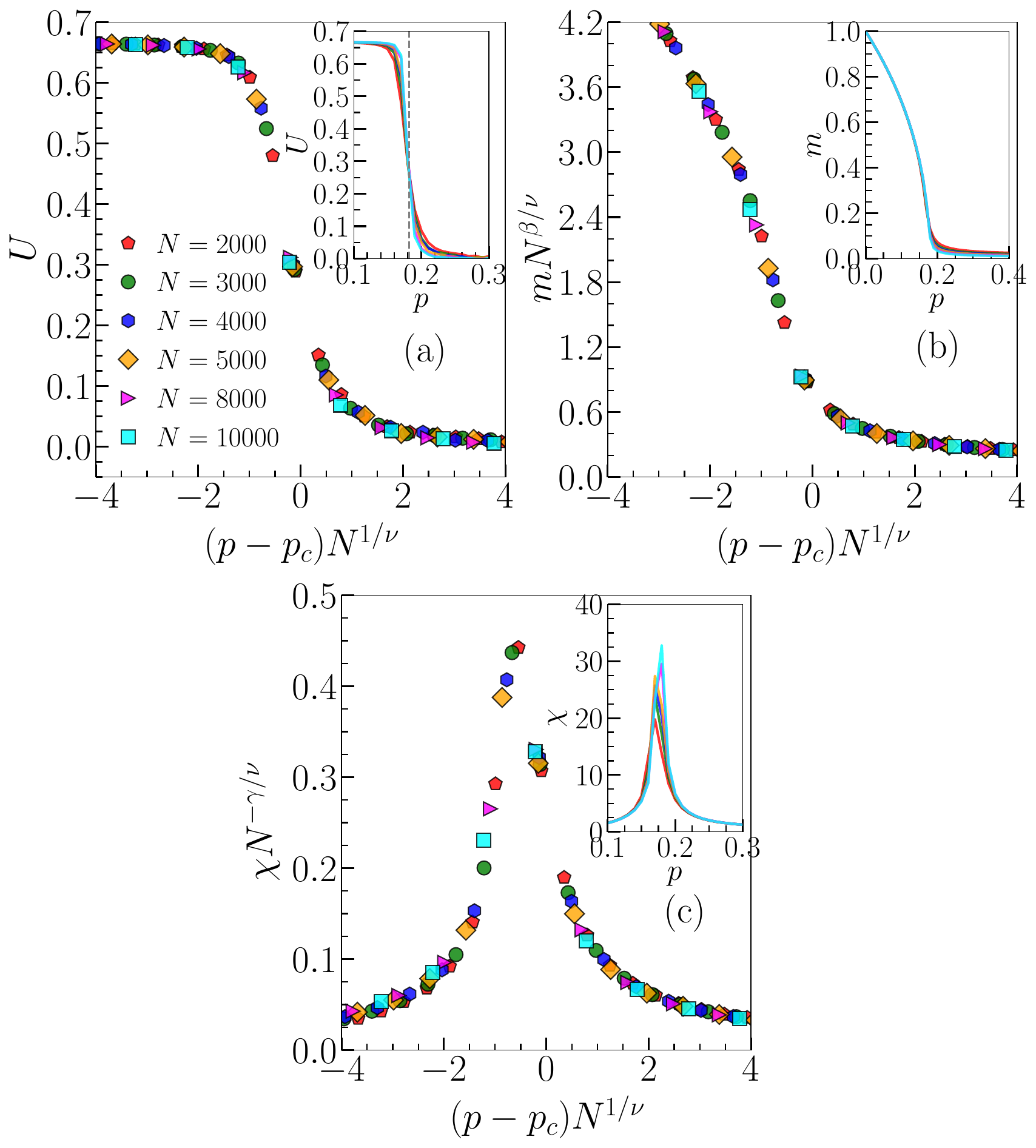}
    \caption{The M-C simulation results of the model on the combined B-A and W-S network for the model with anticonformity for the order parameter \( m \), susceptibility \( \chi \), and Binder cumulant \( U \) across various population sizes \( N \). The model undergoes a continuous phase transition with a critical point at \(  p_c \approx 0.182 \). The best critical exponents of the model are \( \beta \approx 0.5 \), \( \nu \approx 2.0 \), and \( \gamma \approx 1.0 \) . These results suggest that the model is identical to the model on the complete graph and falls into the mean-field Ising universality class.}\label{fig:combined_anti}
\end{figure}

The numerical results for \( m \), \( \chi \), and \( U \) for the model with anticonformity are shown in Fig.~\ref{fig:combined_anti}. This model also undergoes a continuous phase transition with a critical point at \( p_c \approx 0.182 \). The finite-size scaling analysis results show that the model has the same critical exponents as the model with independence: \( \nu \approx 2.0 \), \( \gamma \approx 1.0 \), and \( \beta \approx 0.5 \). This indicates that the model also belongs to the same universality class as the mean-field Ising model. 
Note that the critical exponents for the model on both the complete graph and heterogeneous networks satisfy the hyperscaling relation \(\nu' d_c = 2\beta + \gamma\), indicating the same effective exponent \(\nu' = 1/2\) and critical dimension \(d_c = 4\) as the model on the complete graph, related to the lack of any spatial dimension, analogous to other complex networks \cite{hong2007finite}.

In general, the universality class of the model in these heterogeneous networks requires further verification, given the scope limitations previously established, such as the fixed number of agent interactions, which may reduce the impact of the networks' hub role. Allowing each agent or node to interact with all of its nearest neighbors would be an intriguing point for further analysis, as it could provide more comprehensive insights into the system's critical behavior.

\section{Summary}
\label{sec.4}
This paper examines the impact of ``social noises," characterized by independence and nonconformity behaviors, on the order-disorder phase transition based on the majority rule model. The study encompasses homogeneous networks, including complete graphs and two- and three-dimensional square lattices, and heterogeneous networks like B-A, W-S, and E-R networks. A probability parameter, denoted as \( p \), dictates the likelihood of agents acting independently or nonconformist, while the probability for agents to conform to the majority opinion is \( (1-p) \). The parameter \( p \) is responsible for the occurrence of the phase transition, functioning similarly to temperature in magnetization systems such as the Ising model.

Our results, obtained both analytically (for the complete graph) and through numerical simulations, show that the model undergoes a continuous phase transition across all networks, with different critical points summarized in Table~\ref{tab:table1}. For all networks (the detailed structure of the networks, particularly for complex or heterogeneous networks, can be found in sections \textbf{Model and Methods} and \textbf{Result and Discussion}), the critical point for the model with anticonformity is lower than that for the model with independence. The critical exponents for the models on the homogeneous and heterogeneous networks were obtained through finite-size scaling analysis. Based on the critical exponent data, our results suggest that the model on the complete graph belongs to the same universality class as the mean-field Ising model. The model on the 2-D lattice belongs to the same universality class as the 2-D Ising model, and the model on the 3-D lattice belongs to the 3-D Ising universality class.

Under the scenario of restricting the number of agent interactions, as in the homogeneous networks, the universality class of the model in the heterogeneous networks—such as the B-A network and the combined networks—also aligns with that of the mean-field Ising model. However, since not all agents have the same connectivity or neighbors, further verification of the model’s critical characteristics in these heterogeneous networks is necessary, for example, by allowing each agent to be influenced by their actual number of neighbors and not to restrict them.

\begin{table}[tb]
\centering
\caption{Critical point $p_c$ and critical exponents $\beta, \gamma, \nu$ of the majority rule model on networks.}
\vspace{8pt}
\small
\begin{tabularx}{0.99\linewidth}{@{}lXlllrr@{}}
\hline
\multirow{2}{*}{Networks}  & \multirow{2}{*}{Soc. Noises} & \multirow{2}{*}{$n$} &\multirow{2}{*}{$p_c$}  &  \multicolumn{3}{c}{Critical Exponents}  \\ \cmidrule(l){5-7}
& & & & $\beta$ & $\gamma$ & $\nu$ \\
\hline
\multirow{2}{*}{C.G.} & Indep. & \multirow{2}{*}{$3$} & $1/3$ & $0.5$ & $1.0$ & $2.0$  \\
& Antic. &  & $1/4$  & $0.5$ & $1.0$ & $2.0$  \\
\hline
\multirow{2}{*}{2-D} & Indep. & \multirow{2}{*}{$5$} & $0.106$ & $0.125$ & $1.75$ & $1.0$  \\
& Antic. &  & $0.062$ & $0.125$ & $1.75$ & $1.0$  \\
\hline
\multirow{2}{*}{3-D} & Indep. & \multirow{2}{*}{$7$} & $0.311$ & $0.326$ & $1.237$ & $0.630$  \\
& Antic. &  & $0.268$ & $0.326$ & $1.237$ & $0.630$  \\
\hline
\multirow{1}{*}{B-A} & Indep. & \multirow{1}{*}{$3$} & $0.245$ & $0.5$ & $1.0$ & $2.0$  \\
\hline
\multirow{2}{*}{B-A and W-S} & Indep. & \multirow{2}{*}{$3$} & $0.280$ & $0.5$ & $1.0$ & $2.0$  \\
& Antic. &  & $0.182$ & $0.5$ & $1.0$ & $2.0$  \\
\hline
\end{tabularx} \label{tab:table1}
\end{table}

From a social systems perspective, the phase transition phenomena observed in the model can describe transitions in social dynamics, such as shifts from consensus to polarization. When \( |m| = 1 \), complete consensus occurs in the absence of independent or anticonformist agents. In the presence of these agents, a mix of majority and minority opinions exists below the critical point. However, when \( p \geq p_c \), the system becomes stuck in a stalemate. Comparing the critical points of the independence and anticonformity models reveals that the anticonformity model has a lower critical point than the independence model. These findings suggest that systems with anticonformity agents are more likely to reach a stalemate than those with independent agents.

\section*{Author Contributions}
\textbf{R.~Muslim:} Main contributor, Conceptualization,  Methodology, Writing, Software, Formal analysis, Validation, Visualisation, Review \& editing, Supervision. \textbf{D.~A.~Mulya:} Writing, Software, Formal analysis, Visualisation. \textbf{Z.~Akbar:} Writing, Formal analysis, Validation, Review \& editing.  \textbf{R.~A.~NQZ:} Formal analysis, Validation, Review \& editing. All authors read and reviewed the paper.
\section*{Declaration of Interests}
The contributors declare that they have no apparent competing business or personal connections that might have appeared to have influenced the reported work.
\section*{Acknowledgments}
The authors thank the BRIN Research Center for Quantum Physics for providing the mini High-Performance Computing (HPC) facility for numerical simulations. \textbf{D.~A.~Mulya} expresses sincere gratitude for the support from the Research Assistant program under BRIN talent management, as stipulated in Decree Number 60/II/HK/2023.
\appendix
\section{General formulation for any odd number of agents on the complete graph}
\subsection{\label{appendix1} Model with independence }
For the model with independence on the complete graph, an odd number of agents \( n \) is randomly selected from the population. With a probability \( p \), these \( n \) agents act independently, and with a probability \( 1/2 \), each agent flips its opinion. If the agents do not adopt independent behavior, each agent, with a probability \( 1-p \), will adopt the majority opinion. Assume that each agent has two possible opinions; thus, there are \( 2^n \) possible opinion configurations, with \( 2^n - 2 \) configurations that follow the majority rule and \( 2^{n-1} - 1 \) configurations that will transition to all \( +1 \) or all \( -1 \).

\begin{figure}[tb]
    \centering
    \includegraphics[width = \linewidth]{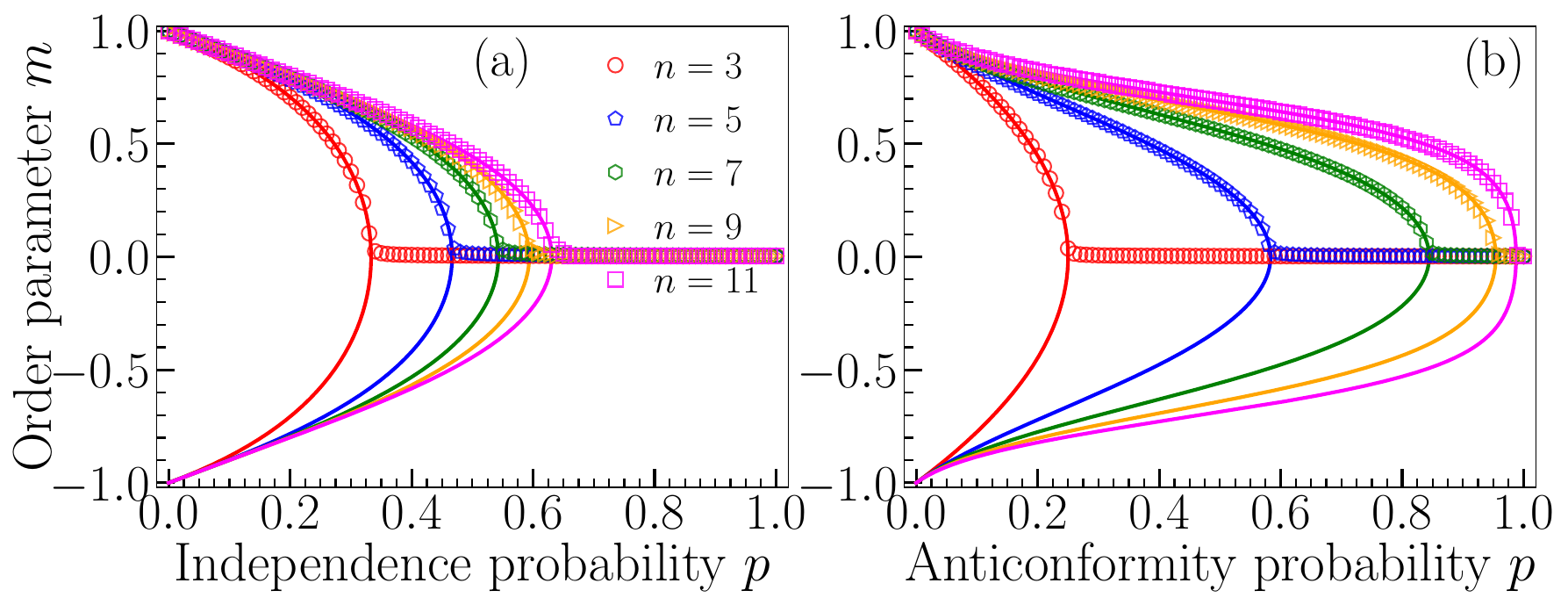}
    \caption{\textcolor{black}{The comparison between the analytical results (solid lines) in Eq.~\eqref{eq:critical_gen1} for the model with independence [panel (a)] and in Eq.~\eqref{eq:critical_gen2} for the model with anticonformity [panel (b)], where \( m = 2 r_{st} - 1 \), and the Monte Carlo simulations (markers) for different group sizes \( n \), shows strong agreement. It can be observed that the model undergoes a continuous phase transition for all group sizes \( n \) at a specific critical independence point \( p_c \). The population size is \( N = 10^5 \), and each data point represents an average over \( 10^5 \) independent realizations.}}\label{fig:critical_indep}
\end{figure}

The probabilities for the independent agents to change their opinions from \( -1 \) to \( +1 \) and from \( +1 \) to \( -1 \) are \( pn\left(1-r\right)/2 \) and \( pnr/2 \), respectively, with \( r \) representing the density of \( +1 \) opinions in the system. Among the \( 2^{n-1} - 1 \) configurations of agents that follow the majority rule, for each configuration of  \( 2^{n-1} - 1 \), half of the agents will adopt the majority opinion. If we define \( k \) as the number of agents with the \( +1 \) opinion, then there are \( n-k \) agents with the \( -1 \) opinion. The number of ways to choose the agents with the \( +1 \) opinion in the system is \( \binom{n}{k} \), and the probabilities of the agents having opinions \( +1 \) and \( -1 \) are \( r^k \) and \( (1-r)^{n-k} \), respectively. Therefore, the density probability of the \( +1 \) opinion increasing in the system can be written as:
  \begin{align}
    \rho^{+}(r) = \frac{np}{2} \left(1 - r\right) + \left(1 - p\right) \sum_{k=(n+1)/2}^{n-1} \left(n-k\right) \binom{n}{k} r^k \left(1 - r\right)^{n-k}, \label{eq:app1}.
\end{align}
In a similar manner, for agents adopting the \( -1 \) opinion, the density probability of the \( -1 \) opinion increasing is:
\begin{equation}
         \rho^{-}(r) = \frac{np}{2} r + \left(1 - p\right) \sum_{k=(n+1)/2}^{n-1} \left(n-k\right) \binom{n}{k} \left(1 - r\right)^k r^{n-k} \label{eq:app2},
\end{equation}
For \( n=3 \), Eqs.~\eqref{eq:app1} and \eqref{eq:app2} reduce to Eqs.~\eqref{eq10} and \eqref{eq11}. For the stationary condition, where \( \rho^{+}(r) = \rho^{-}(r) \), the stationary probability is obtained as follows:
\begin{equation}\label{eq:critical_gen1}
p = \dfrac{l\sum_{k=(n+1)/2}^{n-1} \left(n-k\right)\binom{n}{k} [\left(1-r_{\text{st}}\right)^k r_{\text{st}}^{n-k} - r_{\text{st}}^k \left(1-r_{\text{st}}\right)^{n-k} ]}{1 + l \sum_{k=(n+1)/2}^{n-1} \left(n-k\right) \binom{n}{k} [\left(1-r_{\text{st}}\right)^k r_{\text{st}}^{n-k} - r_{\text{st}}^k \left(1-r_{\text{st}}\right)^{n-k} ]}
\end{equation}
where $l = 2/\left[n \left(1-2r_{\text{st}}\right)\right]$. \textcolor{black}{Based on Eq.~\eqref{eq:critical_gen1}, obtaining an explicit expression for \( r_{st} \) as a function of \( p \) and \( n \) is quite challenging. However, we can directly plot \( p \) versus \( m = 2r_{st} - 1 \) for several values of \( n \) from Eq.~\eqref{eq:critical_gen1} and compare it with the results of numerical simulations to visualize the system's phase diagram, as shown in panel (a) of Fig.~\ref{fig:critical_indep}. It can be observed that the model undergoes a continuous phase transition for all values of \( n \).}

\begin{figure}[tb]
    \centering
    \includegraphics[width = 0.6\linewidth]{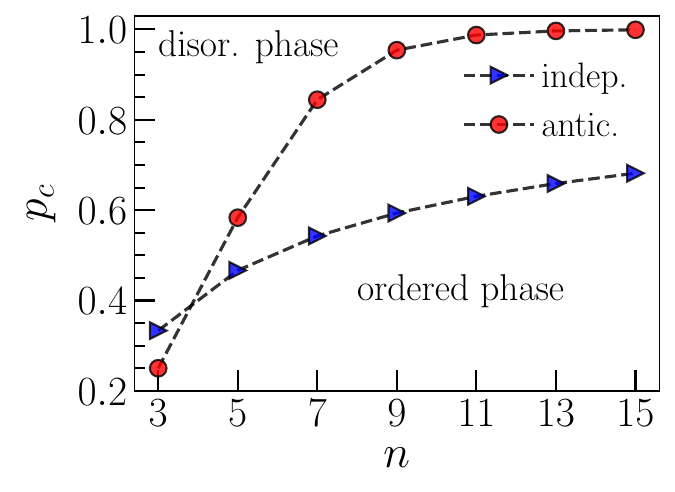}
    \caption{\textcolor{black}{The critical point of the model with independence (blue marker) based on Eq.~\eqref{eq:crit_final1} and with anticonformity (red marker) based on Eq.~\eqref{eq:crit_final2}. The critical point \( p_c \) increases as \( n \) increases. Below the critical point \( p < p_c \), the system is in an ordered phase, while above the critical point \( p \geq p_c \), the system is in a completely disordered phase.}}\label{fig:critical_point}
\end{figure}

By evaluating Eq.~\eqref{eq:critical_gen1} near $r_{st} = 1/2$,  namely $r_{st} = 1/2 +\delta$, where $\delta \ll 1$, we obtain the critical point that makes the model undergo a continuous phase transition for any odd number $n$ is:
\textcolor{black}{\begin{equation} \label{eq:crit_final1}
    p_c(n) =\dfrac{\sum_{k=(n+1)/2}^{n-1} \binom{n}{k} (n-k)(2k-n) }{n \, 2^{n-2} + \sum_{k=(n+1)/2}^{n-1} \binom{n}{k} (n-k)(2k-n) }
\end{equation}}
From Eq.~\eqref{eq:crit_final1}, it can be seen that the critical point \(p_c\) increases with the size of the agents \(n\) as shown in Fig.~\ref{fig:critical_point}.

\subsection{\label{appendix2} Model with anticonformity}
For the model with anticonformity on the complete graph, an odd number of agents \( n \) is randomly selected from the population. With a probability of \( p \), these \( n \) agents act as anticonformists; if these agents hold the same opinion, they will flip their opinions. If the agents do not adopt anticonformist behavior, then each agent, with a probability of \( 1-p \), will adopt the majority opinion. Thus, we can generally formulate the changes in the probability density of opinion \( +1 \) and opinion \( -1 \) for odd \( n \) as follows:
  \begin{align}
    \rho^{+}(r) & = np\left(1 - r\right)^{n} + \left(1 - p\right) \sum_{k=(n+1)/2}^{n-1} \left(n-k\right) \binom{n}{k} r^k \left(1 - r\right)^{n-k}, \label{eq:app3}\\
     \rho^{-}(r) & = npr^{n} + \left(1 - p\right) \sum_{k=(n+1)/2}^{n-1} \left(n-k\right) \binom{n}{k} \left(1 - r\right)^k r^{n-k} \label{eq:app4}.
\end{align}
For \( n = 3 \), Eqs.~\eqref{eq:app3} and \eqref{eq:app4} reduce to Eqs.~\eqref{eq12} and \eqref{eq13}. In general, the stationary probability for this model is:
\begin{equation}\label{eq:critical_gen2}
    p = \dfrac{\sum_{k=(n+1)/2}^{n-1} \binom{n}{k} \left(n-k\right) [\left(1-r_{\text{st}}\right)^k r_{\text{st}}^{n-k} - r_{\text{st}}^k \left(1-r_{\text{st}}\right)^{n-k}]}{l' + \sum_{k=(n+1)/2}^{n-1} \left(n-k\right) \binom{n}{k} [\left(1-r_{\text{st}}\right)^k r_{\text{st}}^{n-k} - r_{\text{st}}^k \left(1-r_{\text{st}}\right)^{n-k}]},
\end{equation}
where $l' = n\left[\left(1-r_{\text{st}}\right)^n - r_{\text{st}}^n\right]$. \textcolor{black}{The plot of Eq.~\eqref{eq:critical_gen2} for several values of \( n \) is shown in the panel (b) of Fig.~\ref{fig:critical_indep}. It can be observed that the model undergoes a continuous phase transition for all values of \( n \).}

\begin{figure}[tb]
    \centering
    \includegraphics[width = 0.65\linewidth]{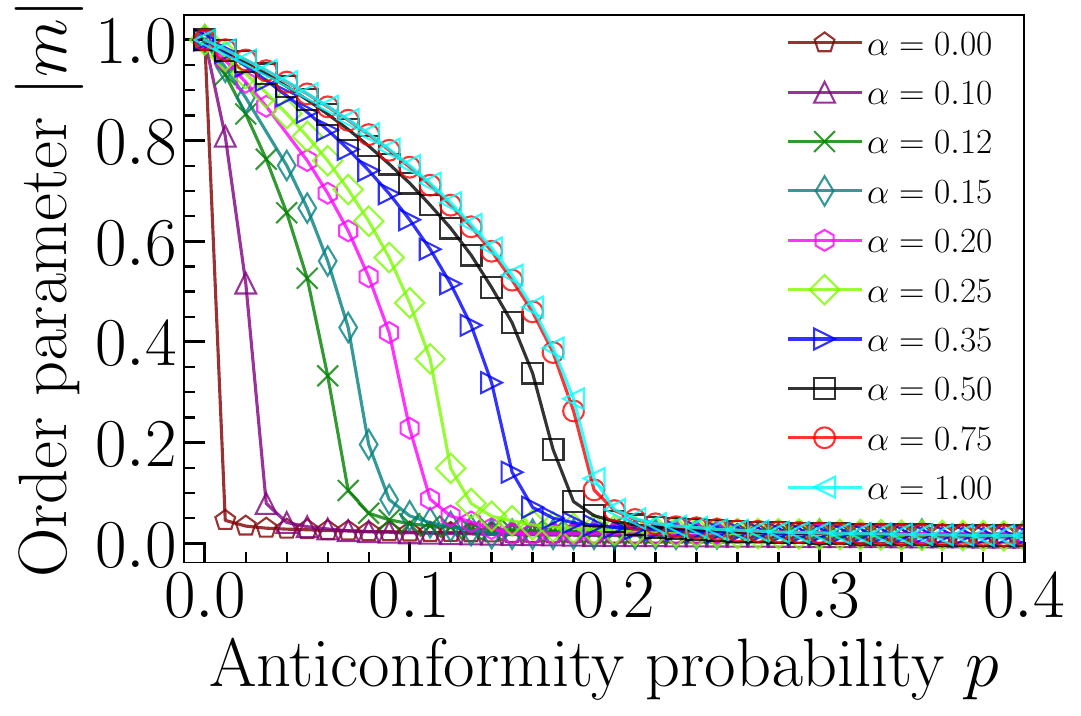}
    \caption{\textcolor{black}{The M-C simulation results for the model with anticonformity on the W-S network, with various values of rewiring probability \( \alpha \) and an average degree of \( \langle k \rangle = 4 \), are presented. It can be observed that the critical point of the model increases as \( \alpha \) increases. For \( \alpha = 0.75 \) and \( \alpha = 1.0 \), there is a slight overlap, but some differences are still noticeable. The population size is \( N = 10^4 \), and each data point represents the average of \( 10^5 \) independent realizations.}}\label{fig:var_beta}
\end{figure}

\textcolor{black}{Similarly, for the model with independence, by evaluating Eq.~\eqref{eq:critical_gen2} near \( r_{st} = 1/2 \), we obtain the critical point at which the model undergoes a continuous phase transition for any odd number \( n \) as:}
\textcolor{black}{\begin{equation}\label{eq:crit_final2}
    p_c(n) = \dfrac{\sum_{k=(n+1)/2}^{n-1} \binom{n}{k} (n-k)(2k-n)}{n^2 + \sum_{k=(n+1)/2}^{n-1} \binom{n}{k} (n-k)(2k-n)}
\end{equation}}
Based on Eq.~\eqref{eq:crit_final2}, the critical point of the system also increases with the size of agents \(n\), as shown in Fig.~\ref{fig:critical_point}.
\textcolor{black}{\section{Some other numerical simulation results}
\subsection{\label{app:b.1} Order parameter of the model with varying rewiring probabilities \( \alpha \)}
We further analyzed the impact of the rewiring probability $\alpha \in [0, 1]$ on the critical point of the model with anticonformity on the W-S network. The numerical results of the model for the order parameter \( m \) with typical values of the rewiring probabilities \( \alpha \) are exhibited in Fig.~\ref{fig:var_beta}. Here, we examine a group of three agents interacting based on anticonformity behavior described in the model's algorithm. As previously known, no phase transition occurs when $\alpha = 0.00$, as in this case, the W-S network behaves like a 1-D lattice. For $\alpha > 0$, the model undergoes a continuous phase transition, with the critical point increasing as the value of $\alpha$ increases. This phenomenon has also been discussed in Ref.~\cite{campos2003small}.}
\textcolor{black}{\subsection{Critical exponents of the model on the W-S network for $n = 9$}
We further examined the critical exponents of the model with independence for an agent group size \( n = 9 \) on the W-S network with an average node degree \( \langle k \rangle = 10 \) and a rewiring probability \( \alpha = 0.5 \). At this value of \( \alpha = 0.5 \), the network can capture the characteristics of a small-world network, such as a high clustering coefficient and short path lengths across the network \cite{watts1998collective, newman2003structure, barrat2008dynamical}.  In this model, at each Monte Carlo step (time), a node or agent is randomly selected, followed by the random selection of eight neighbors from its total \( k \) neighbors. These nine agents then interact according to the model's algorithm. With probability \( p \), all nine agents adopt anticonformity behavior. If all nine agents share the same opinion or state, they switch to the opposite opinion. If they do not adopt anticonformity behavior, with probability \( 1-p \), the nine agents adopt the majority opinion among them.
}

\begin{figure}[tb]
    \centering
    \includegraphics[width = \linewidth]{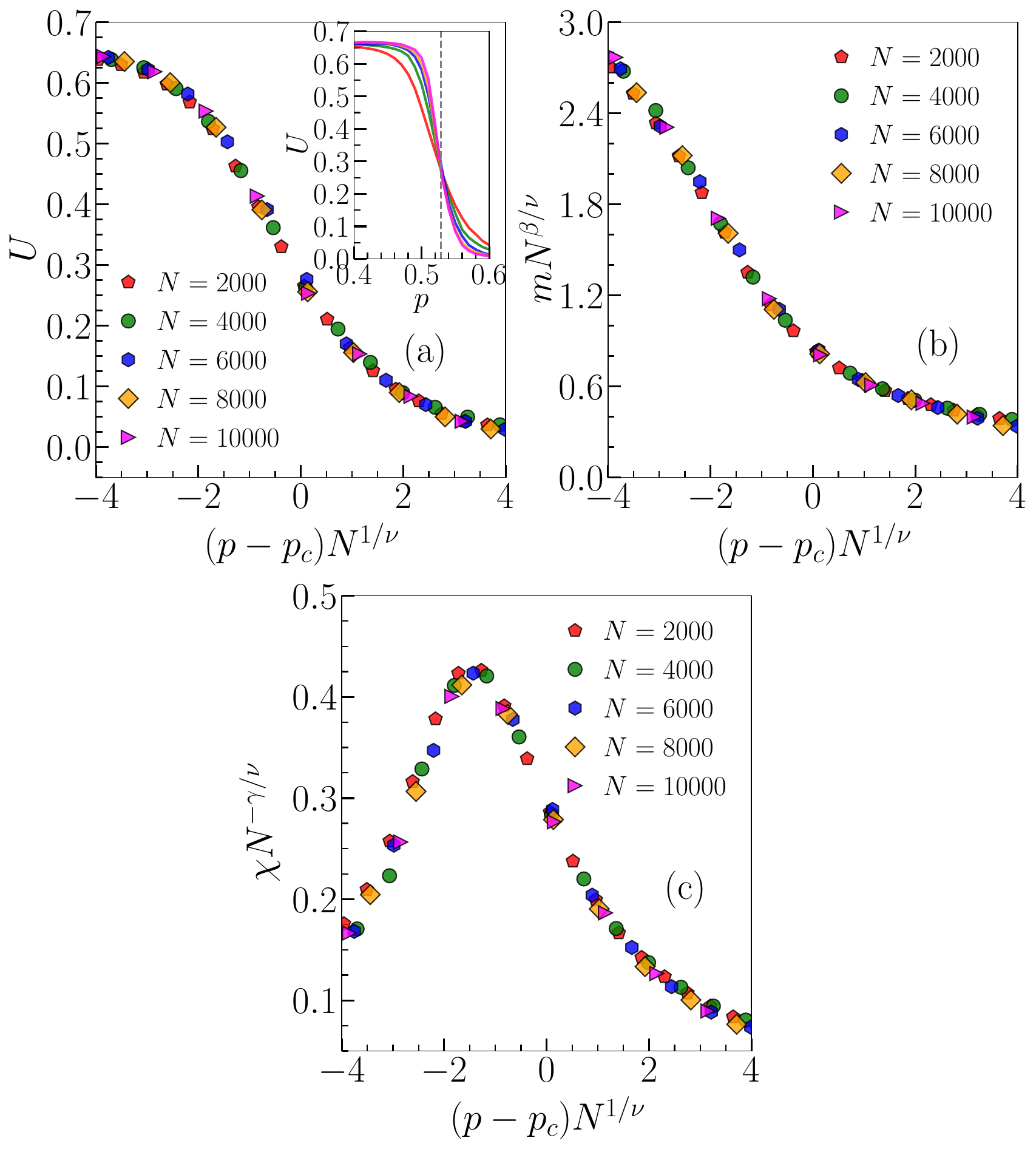}
    \caption{\textcolor{black}{The M-C simulation results of the model on the W-S network for the model with independence for $n = 9$. The model undergoes a continuous phase transition with a critical point at \(  p_c \approx 0.528 \). The best critical exponents of the model are \( \beta \approx 0.5 \), \( \nu \approx 2.0 \), and \( \gamma \approx 1.0 \). Each data point represents the average of \( 5 \times 10^5 \) independent realizations.}}\label{fig:fig_cr_ws}
\end{figure}

\textcolor{black}{
It is shown in Fig.~\ref{fig:fig_cr_ws} that the model undergoes a continuous phase transition at the critical point \( p_c \approx 0.528 \) [see the inset of panel (a)]. Using finite-size scaling on Eqs.~\eqref{eq1} - \eqref{eq2}, the best critical exponents that collapse the data around the critical point \( p_c \) are found to be \( \beta \approx 0.5 \), \( \gamma \approx 1.0 \), and \( \nu \approx 2.0 \). These values of the critical exponents are the same as those for the model with \( n=3 \), reinforcing our conclusion that the model belongs to the same universality class as the mean-field Ising model.}
\textcolor{black}{\subsection{Critical exponent of the model on the multilayer network for $n = 9$}}
\textcolor{black}{We further examined the critical exponents of the model with independence for an agent group size of \( n = 9 \) on the multilayer network, with an average node degree \( \langle k \rangle = 18 \) for the B-A network and \( \langle k \rangle = 10  \) and a rewiring probability \( \alpha = 0.25 \) for the W-S network. A total of \( N_c = N_i/2 \) nodes are interconnected between layers, where \( N_i \) represents the number of nodes in each layer. In this model, a node or agent is randomly selected at each Monte Carlo step (time), followed by the random selection of eight neighbors from its total \( k \) neighbors. These nine agents then interact according to the model's algorithm.
}

\begin{figure}[tb]
    \centering
    \includegraphics[width = \linewidth]{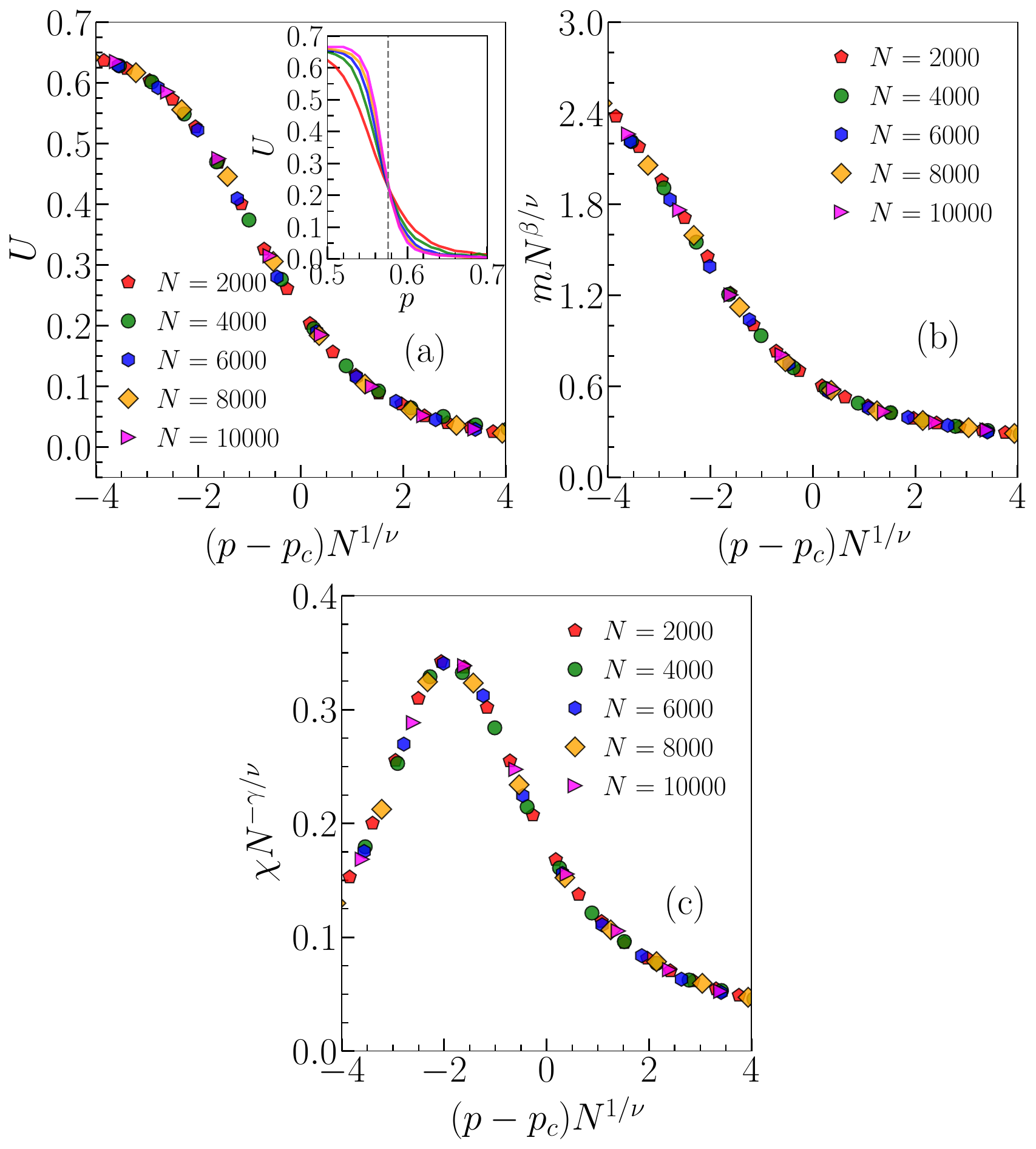}
    \caption{\textcolor{black}{The M-C simulation results for the model with independence on the multilayer networks with \( n = 9 \) show that the model undergoes a continuous phase transition with a critical point at \( p_c \approx 0.565 \). The best critical exponents of the model are \( \beta \approx 0.5 \), \( \nu \approx 2.0 \), and \( \gamma \approx 1.0 \).  Each data point represents the average of \( 5 \times 10^5 \) independent realizations.}}\label{fig:fig_cr_multi}
\end{figure}

\textcolor{black}{
Our results in Fig.~\ref{fig:fig_cr_multi} demonstrate that the model undergoes a continuous phase transition at the critical point \( p_c \approx 0.576 \) [see the inset of panel (a)]. Using finite-size scaling on Eqs.~\eqref{eq1} - \eqref{eq2}, the best critical exponents that collapse the data around the critical point \( p_c \) are found to be \( \beta \approx 0.5 \), \( \gamma \approx 1.0 \), and \( \nu \approx 2.0 \), indicating that the model belongs to the same universality class as the mean-field Ising model.}
\bibliographystyle{elsarticle-num} 

\end{document}